\documentclass[showpacs,aps,prd,nofootinbib,floatfix,amsmath,amssymb,twocolumn]{revtex4}
\usepackage{mathrsfs}
\usepackage{graphicx}
\usepackage[dvipsnames]{xcolor}
\usepackage{dsfont}
\begin{document}

\makeatletter
\newbox\slashbox \setbox\slashbox=\hbox{$/$}
\newbox\Slashbox \setbox\Slashbox=\hbox{\large$/$}
\def\pFMslash#1{\setbox\@tempboxa=\hbox{$#1$}
  \@tempdima=0.5\wd\slashbox \advance\@tempdima 0.5\wd\@tempboxa
  \copy\slashbox \kern-\@tempdima \box\@tempboxa}
\def\pFMSlash#1{\setbox\@tempboxa=\hbox{$#1$}
  \@tempdima=0.5\wd\Slashbox \advance\@tempdima 0.5\wd\@tempboxa
  \copy\Slashbox \kern-\@tempdima \box\@tempboxa}
\def\FMslash{\protect\pFMslash}
\def\FMSlash{\protect\pFMSlash}
\def\miss#1{\ifmmode{/\mkern-11mu #1}\else{${/\mkern-11mu #1}$}\fi}
\makeatother

\title{Lorentz violation in nucleon electromagnetic moments}

\author{Javier Monta\~no-Dom\'inguez$^{(a)}$}\author{H\'ector Novales-S\'anchez$^{(b)}$}
\author{M\'onica Salinas$^{(b)}$}
\author{J. Jes\'us Toscano$^{(b)}$}
\affiliation{$^{(a)}$C\'atedras Conacyt-Facultad de Ciencias F\'isico Matem\'aticas,
Universidad Michoacana de San Nicol\'as de Hidalgo,
Av. Francisco J. M\'ugica s/n, C. P. 58060, Morelia, Michoac\'an, M\'exico.\\
$^{(b)}$Facultad de Ciencias F\'isico Matem\'aticas, Benem\'erita Universidad Aut\'onoma de Puebla, Apartado Postal 1152 Puebla, Puebla, M\'exico}

\begin{abstract}
Yukawa couplings in the Lorentz- and $CPT$-violating Standard Model Extension (SME) induce quantum contributions to the electromagnetic moments of quarks, which are calculated in the present paper and then used to define contributions to electromagnetic moments of nucleons. High-sensitivity measurements of the proton and neutron electromagnetic moments then yield constrains on SME coefficients, reaching bounds as restrictive as $10^{-12}$. This approach notably grants access to SME effects from the second and third quark families.
\end{abstract}


\maketitle

\section{Introduction}
Symmetries are a main aspect behind the construction of sensible field theories that pursue a faithful description of nature. Most formulations of physics beyond the Standard Model (SM) bear Lorentz invariance, which has not been observed to fail so far, though high-energy descriptions motivating the existence of Lorentz-symmetry nonconservation exist~\cite{KoSa,KoPo,KoPo2,CHKLO}. The lack of experimental evidence pointing towards the actual existence of Lorentz violation and its presumable origin favors the usage of effective Lagrangians, which provide model-independent descriptions of effects at the reach of current experimental sensitivity~\cite{Wudka}, but which come from high-energy field-theory formulations. Built upon the generality provided by effective theories, the Lorentz- and $CPT$-violating SM Extension (SME)~\cite{CoKo1,CoKo2}, by Colladay and Kosteleck\'y, is a valuable tool for phenomenologists to look for traces, even if tiny, of effects produced by the breaking of Lorentz invariance. Lorentz violation in the SME Lagrangian is characterized by constant coefficients that transform as tensors with respect to observer Lorentz transformations, but are, contrastingly, invariant under particle Lorentz transformations~\cite{CoKo1,CoKo2}, thus defining preferred directions in spacetime. \\

A plethora of papers has been dedicated to bound Lorentz-violation coefficients, among the huge list of parameters characterizing the sectors of the SME. A comprehensive and annually-updated list of constraints on the coefficients of this effective-Lagrangian description, taken from a variety of papers by different authors, is provided in Ref.~\cite{KoRu}. The present work shows a calculation of contributions from the Yukawa sector of the renormalizable-SME~\cite{CoKo2,HMNSTV}, usually dubbed minimal SME, to the anomalous magnetic moment (AMM) and to the electric dipole moment (EDM) of quarks, as well as an estimation of bounds on SME coefficients of this sector, in accordance with current bounds on the neutron and the proton electromagnetic moments (EMM)~\cite{pMM,nEDM,PDG}. These new-physics contributions to quarks EMMs are extracted from the expression of the quark electromagnetic vertex $A_\mu q_Aq_A$, with $q$ denoting some $A$-flavored quark and $A_\mu$ refering to the electromagnetic field, that arises from leading one-loop Feynman diagrams generated by the minimal-SME Yukawa-sector couplings. The assumption of Lorentz nonconservation for the calculation of the vertex $A_\mu q_Aq_A$ enlarges the usual Lorentz-invariant parametrization~\cite{HIRSS,NPR}, which, nevertheless, still includes the usual Lorentz-preserving EMM form factors. Therefore, SME coefficients participating in these EMM contributions have all their spacetime indices fully contracted among themselves and are found to emerge, for the first time, at the second order in Lorentz-violation. \\

Our calculation has been carried out by following a perturbative approach in which 2-point Lorentz-violation insertions are building blocks of Feynman diagrams contributing to $A_\mu q_Aq_A$. While, in practice, dominant contributions, found to emerge from Schwinger-like diagrams, were the ones used to constrain SME coefficients, the calculation of all the contributing diagrams has been performed. Gauge-dependent diagrams have been calculated in the unitary gauge, thus reducing the number of involved diagrams. Even though a practical drawback of this gauge choice is the increment of the superficial degree of divergence of loop integrals, all contributions were found to be ultraviolet (UV) finite. On the other hand, Another type of divergences are the infrared (IR) ones,
which we find to remain within our expressions for both the AMM and the EDM form factors, although such divergences are expected to vanish from cross sections. While this means that the resulting form factors are not to be understood as observables, IR divergences have been, on these grounds, disregarded in order to estimate these quantities, as they will be present in physical processes. Since SME Yukawa couplings are quark-flavor changing, the SME contributions to quark EMMs yield bounds on coefficients associated to the second and third quark families, despite restrictions are established from EMMs of nucleons, exclusively constituted by first-family quarks.
\\

The remainder of the paper goes as follows: in Sec.~\ref{theory}, the SME Yukawa terms, required for the calculation, are provided; a discussion on the calculation of contributions to the quark electromagnetic vertex and the electromagnetic moments is given in Sec.~\ref{calculations}; bounds on SME coefficients are estimated in Sec.~\ref{estimations}; finally, in Sec.~\ref{conclusions} conclusions are presented.

\section{The minimal-SME Yukawa sector}
\label{theory}
First given in Ref~\cite{CoKo2}, the quark Yukawa sector of the minimal SME reads
\begin{eqnarray}
& \displaystyle
{\cal L}^{\rm SME}_{{\rm Y},q}=-\frac{1}{2}(H_u)^{AB}_{\mu\nu}\overline{Q_A}\,\tilde{\phi}\,\sigma^{\mu\nu}u'_B+{\rm H.c.}
\nonumber \\ & \displaystyle
\hspace{1.35cm}
-\frac{1}{2}(H_d)^{AB}_{\mu\nu}\overline{Q_A}\,\phi\,\sigma^{\mu\nu}d'_B+{\rm H.c.}
\label{firstLVYukawa}
\end{eqnarray}
It involves the ${\rm SU}(2)_L$ spinor doublets $Q_A$ and singlets $u'_A$, $d'_A$, where $A=u,c,t$ is a quark-flavor index. The Higgs doublet, denoted as $\phi$, plays a role as well. These fields couple through Yukawa-like constants $(H_u)^{AB}_{\mu\nu}$ and $(H_d)^{AB}_{\mu\nu}$ bearing both quark-flavor indices $A,B$ and spacetime indices $\mu,\nu$. Under observer Lorentz transformations, $(H_u)^{AB}_{\mu\nu}$ and $(H_d)^{AB}_{\mu\nu}$ transform as 2-tensors, whereas they behave as scalars with respect to particle Lorentz transformations, thus yielding Lorentz violation in Eq.~(\ref{firstLVYukawa}). Moreover, these SME Yukawa-like constants are antisymmetric with respect to their spacetime indices, that is, $(H_f)^{AB}_{\mu\nu}=-(H_f)^{AB}_{\nu\mu}$. Implementation of unitary transformations $U^f_L$ and $U^f_R$, after spontaneous breaking of the electroweak symmetry, then yields
\begin{equation}
{\cal L}_{{\rm Y},\,q}^{\rm SME}=-\frac{1}{2}(v+H)\sum_{f=u,d}\overline{f_A}\Big[(Y_f)^{AB}_{\mu\nu}P_L+(Y_f)^{BA*}_{\mu\nu}P_R\Big]\sigma^{\mu\nu}f_B,
\label{Yukawasector}
\end{equation}
in the unitary gauge. Note that quark-flavor indices $A,B$ in Eq.~(\ref{Yukawasector}) run over either $u,c,t$ or $d,s,b$, depending on whether $f=u$ or $f=d$. Let us emphasize that the Lorentz-violating couplings of the Higgs boson, $H$, to quarks turned out to be quark-flavor changing. Furthermore, $(Y_f)_{\mu\nu}=U_L^{f\dag}(H_f)_{\mu\nu}U^f_R$, so that spacetime-group $4\times4$ matrices $(Y_f)^{AB}$ are antisymmetric. Inspiration is then taken from the relations linking the electromagnetic tensor $F_{\mu\nu}$ with the electric and magnetic fields to define the complex electric-like vector ${\bf e}^{AB}$ and the complex magnetic-like vector ${\bf b}^{AB}$ by 
\begin{equation}
Y^{AB}_{0i}= e^{AB}_{i}, \hspace{0.4cm}
Y^{AB}_{ ij}=\varepsilon_{ijk}b^{AB \,k},
\end{equation}
respectively. \\

\section{SME contributions to quark electromagnetic form factors}
\label{calculations}
In this section, we discuss the calculation of contributions from the Yukawa sector of the minimal SME to the AMMs and EDMs of quarks. Aiming at the accomplishment of such a task, we have followed the tensor-reduction method, by Passarino and Veltman~\cite{PaVe}, relying on computational tools, namely, the software {\it Mathematica}, by Wolfram, along with the packages {\it Feyncalc}~\cite{Feyncalc1,Feyncalc2,Feyncalc3} and {\it Package X}~\cite{PackageX}. The EMMs of fermions are generated exclusively at the loop level, being identified from the general Lorentz-invariant parametrization
\begin{eqnarray}
\Gamma^f_\mu(q^2)&=&ie\big( \gamma_\mu\big( V_f(q^2)-A_f(q^2)\gamma_5 \big)
\nonumber \\ 
&&+\sigma_{\mu\nu}q^\nu\Big( \frac{i}{2m}a_f(q^2)-\frac{1}{e}d_f(q^2)\gamma_5 \Big) \big),
\label{linvpar}
\end{eqnarray}
of the electromagnetic vertex $A_\mu ff$~\cite{HIRSS,NPR}, with $f$ representing some fermion, taken on shell, and $q$ denoting the incoming momentum of the photon, assumed to be off the mass shell. Then the on-shell quantities $a_f(q^2=0)$ and $d_f(q^2=0)$ define the AMM and the EDM, respectively. Calculations of the SM contributions to the electron and muon AMMs, as well as measurements of these quantities, have reached remarkable precision~\cite{eAMMth,muAMMth,muAMMexp,eAMMexp1,eAMMexp2}, rendering the tiny experiment-theory difference a standard place to look for new-physics traces. On the side of the quarks, the authors of Ref.~\cite{BBGHLMR} carried out an estimation of SM contributions to the AMMs of the top and bottom quarks. EDMs are important because of their connection with the interesting phenomenon of $CP$ violation, relevant in order to explain baryonic asymmetry~\cite{Sakharov}. However, EDMs of elementary particles have never been measured and, moreover, contributions from the SM to these quantities are produced for the first time at the three-loop level~\cite{CzKr}, thus being quite suppressed and far beyond the reach of experimental sensitivity. \\

Tree-level vertices $Hq_Aq_B$ are given by the expression for the Yukawa-sector Lagrangian in Eq.~(\ref{Yukawasector}), which also carries terms quadratic in quark fields, used in the present investigation to define bilinear insertions that enter Feynman diagrams. At one loop, there are contributions to the electromagnetic vertex $A_\mu q_Aq_A$ from Feynman diagrams with such 2-point and/or 3-point Lorentz-violating insertions. Note that both the bilinear insertions and the trilinear vertices can induce quark-flavor change in Feynman diagrams. While this feature plays a main role in our calculation, through virtual-particle effects, keep in mind that the present work focuses on diagonal EMMs rather than transition EMMs, so the quark flavors of the external quark fields coincide with each other. The assumption of Lorentz-invariance violation enriches the analytical structure of the parametrization of the electromagnetic vertex. Even so, the resulting parametrization still includes the Lorentz-invariant form factors that characterize Eq.~(\ref{linvpar}), so keep in mind that the quark AMM and the quark EDM are Lorentz-invariant, which means that all SME coefficients within the contributions from the Yukawa sector to these quantities must have all their spacetime indices contracted among themselves, as this ensures invariance of the EMMs under both observer transformations and particle transformations. Note that the SME coefficients under consideration are all antisymmetric in their spacetime indices, meaning that at the first order in Lorentz-violation SME coefficients within such factors, which necessarily appear as traces over Lorentz-indices, vanish, thus forbidding the occurrence of SME contributions to EMMs at the first order in Lorentz-violation coefficients. Therefore, contributions from the minimal-SME Yukawa sector to AMMs and EDMs emerge for the first time from one-loop diagrams at the second order in Lorentz-violating coefficients~\cite{AMNST}. \\

\begin{figure}[ht]
\center
\includegraphics[width=3.2cm]{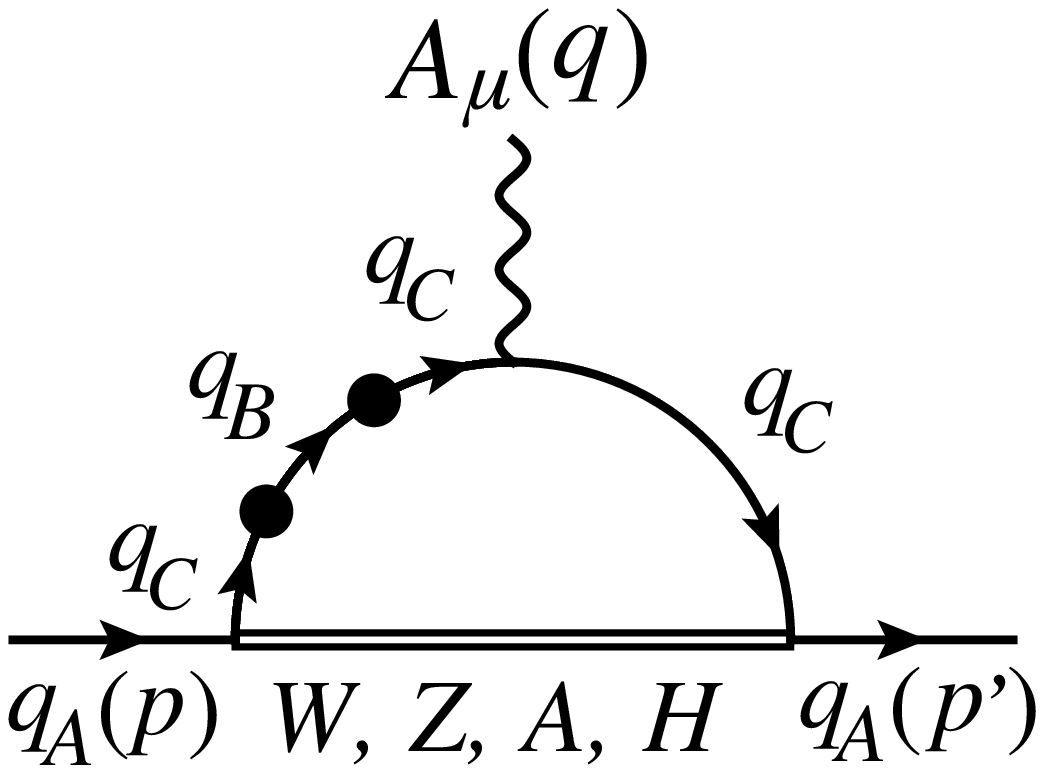}
\hspace{0.3cm}
\includegraphics[width=3.4cm]{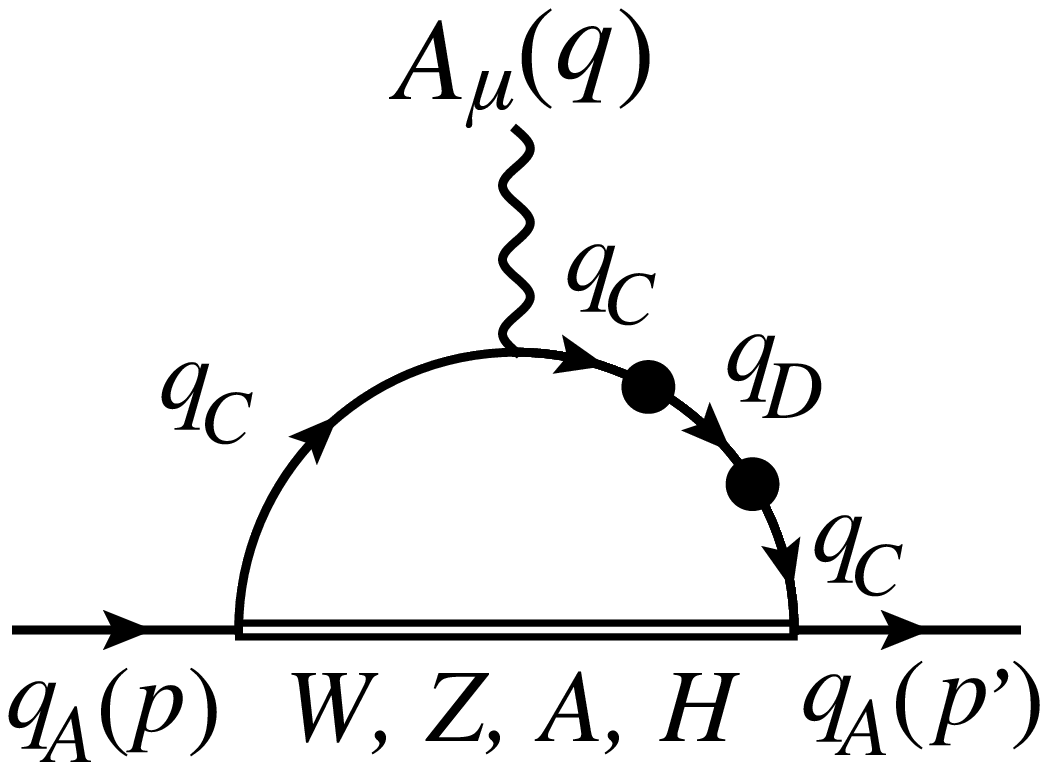}
\includegraphics[width=3.2cm]{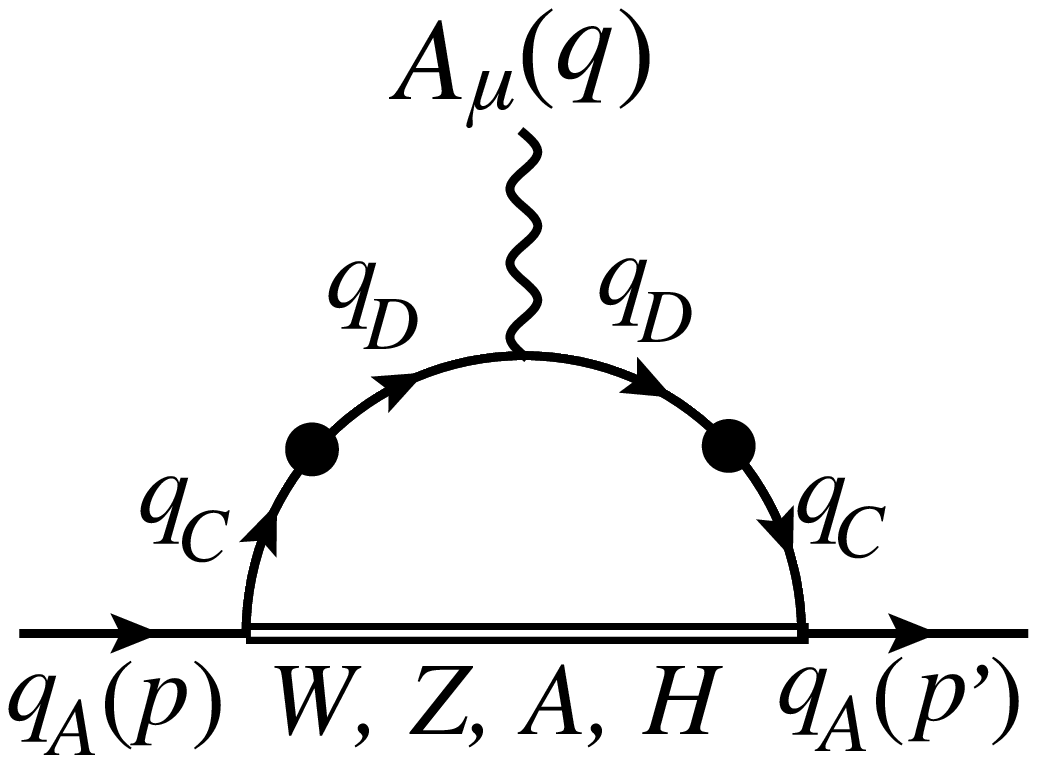}
\hspace{0.3cm}
\includegraphics[width=3.9cm]{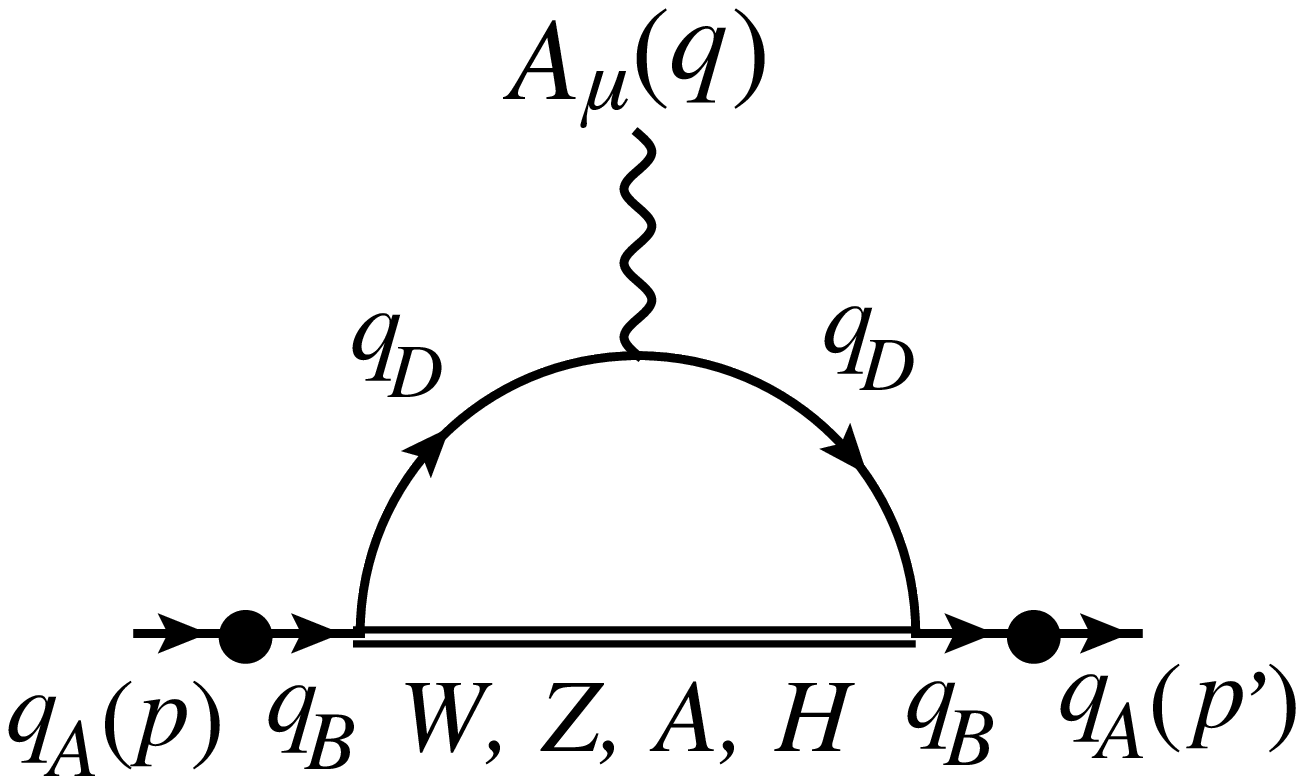}
\includegraphics[width=3.6cm]{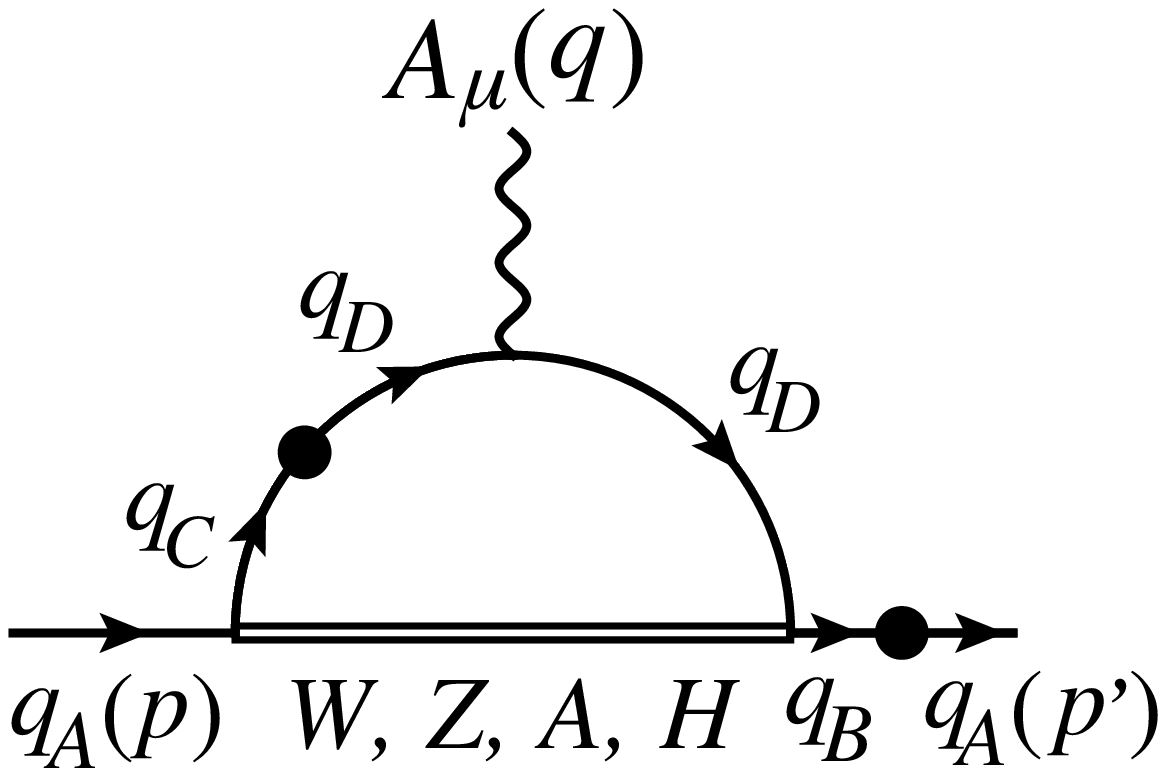}
\hspace{0.3cm}
\includegraphics[width=3.6cm]{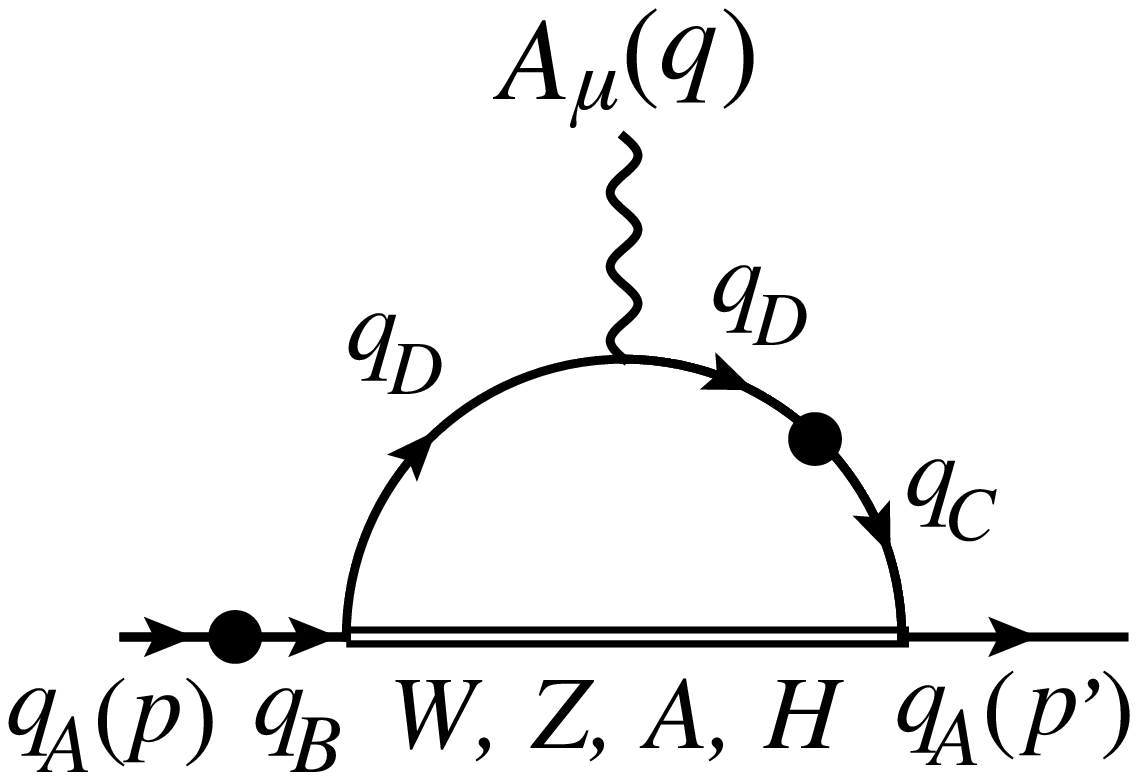}
\includegraphics[width=3.6cm]{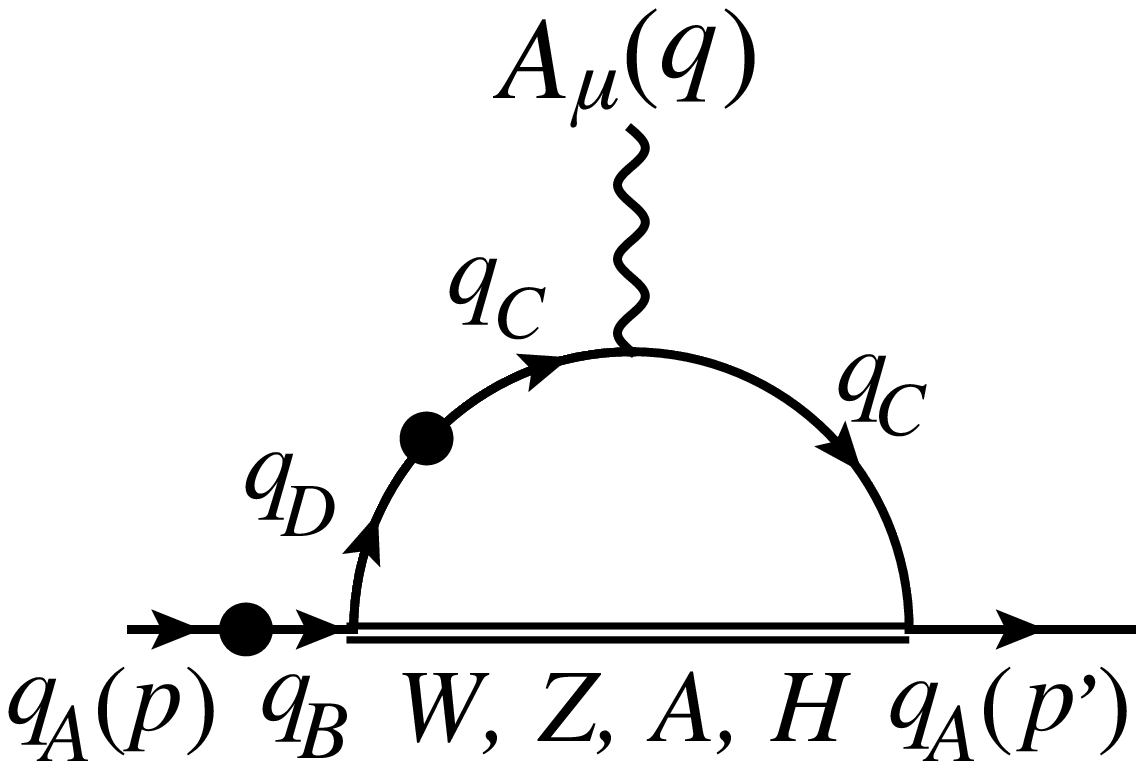}
\hspace{0.3cm}
\includegraphics[width=3.7cm]{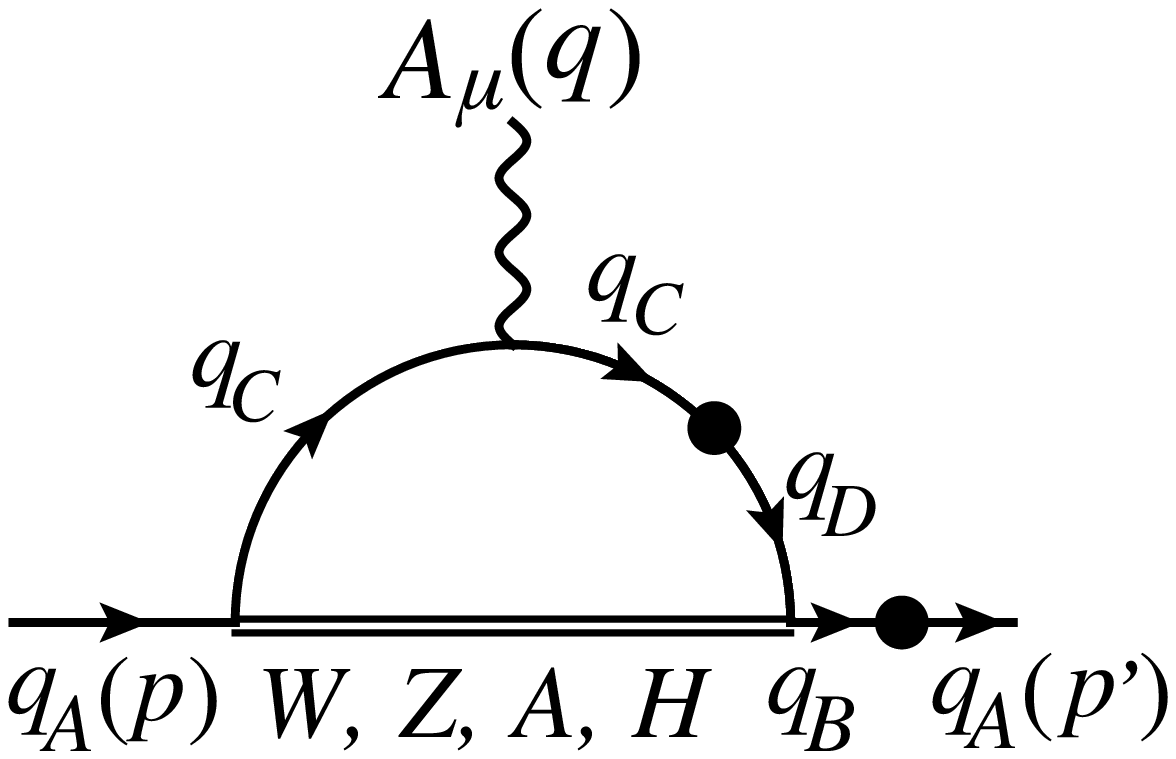}
\caption{\label{gbdiagrams} Feynman diagrams $A_\mu q_Aq_A$ contributing to EMMs, with SME effects entering exclusively through bilinear insertions $q_Aq_B$. If the virtual boson is $Z$, $A$, or $H$, then $C=A$ and $D=B$, while these indices can differ in the case of the $W$ boson, as there are two sources of quark-flavor change.}
\end{figure}

\begin{figure}[ht]
\center
\includegraphics[width=3.2cm]{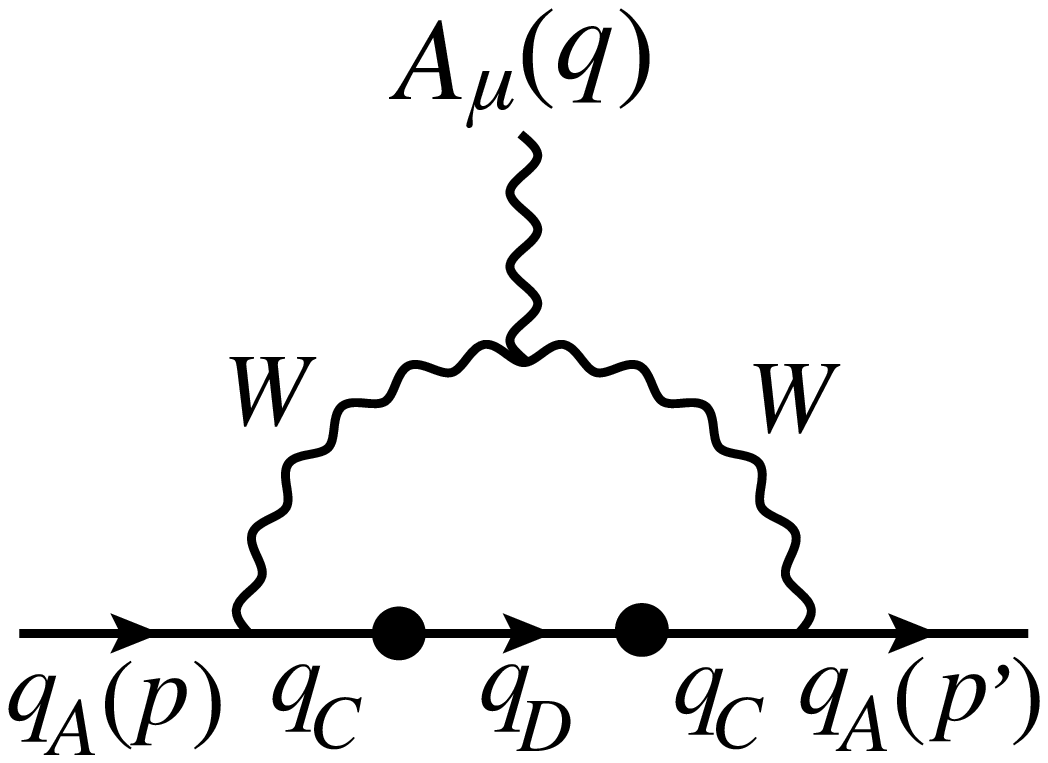}
\hspace{0.3cm}
\includegraphics[width=4.2cm]{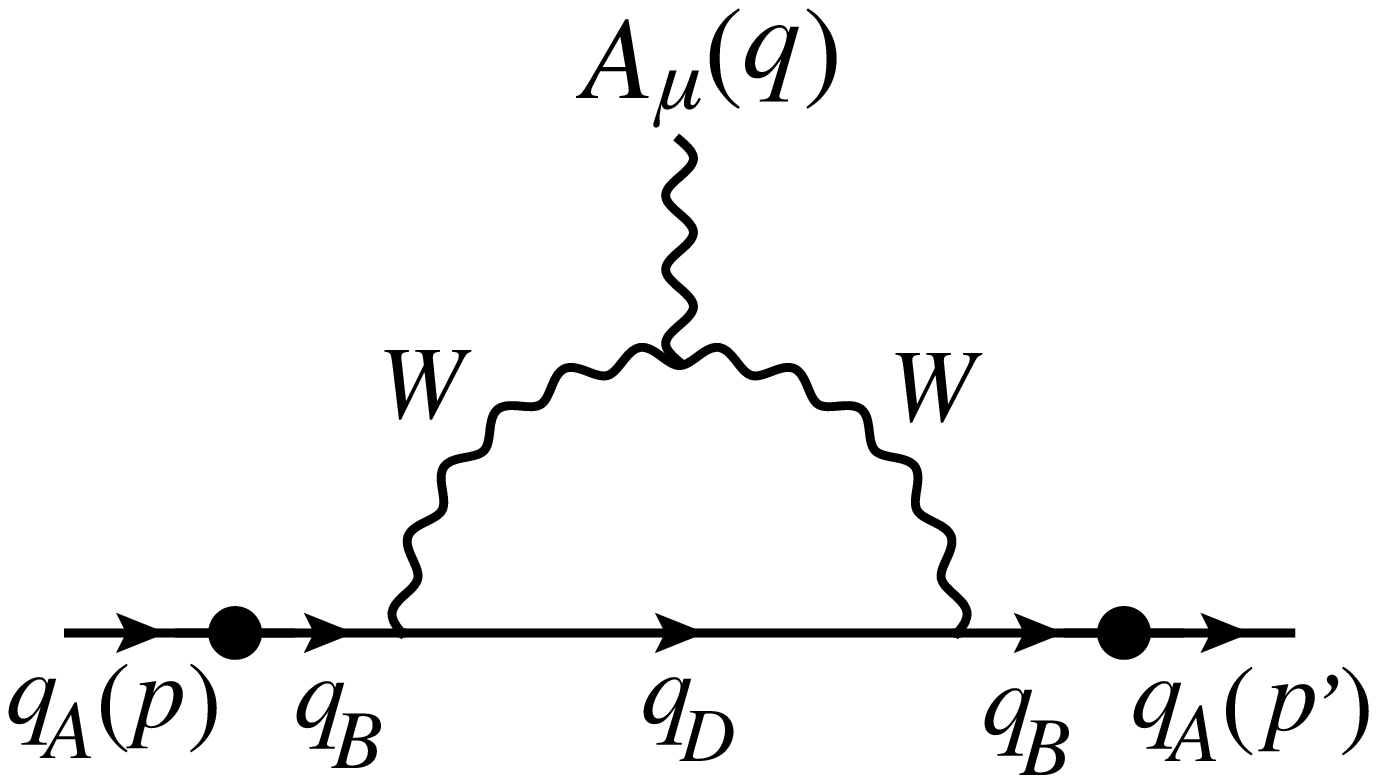}
\includegraphics[width=3.8cm]{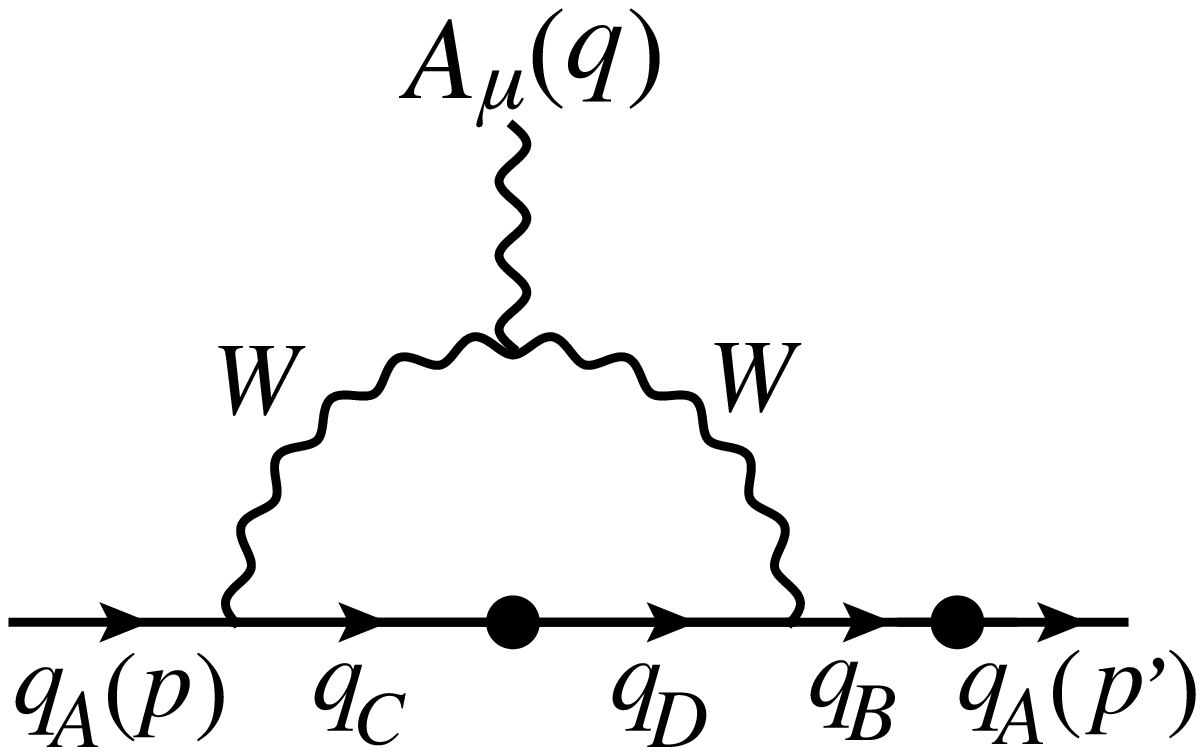}
\hspace{0.3cm}
\includegraphics[width=3.8cm]{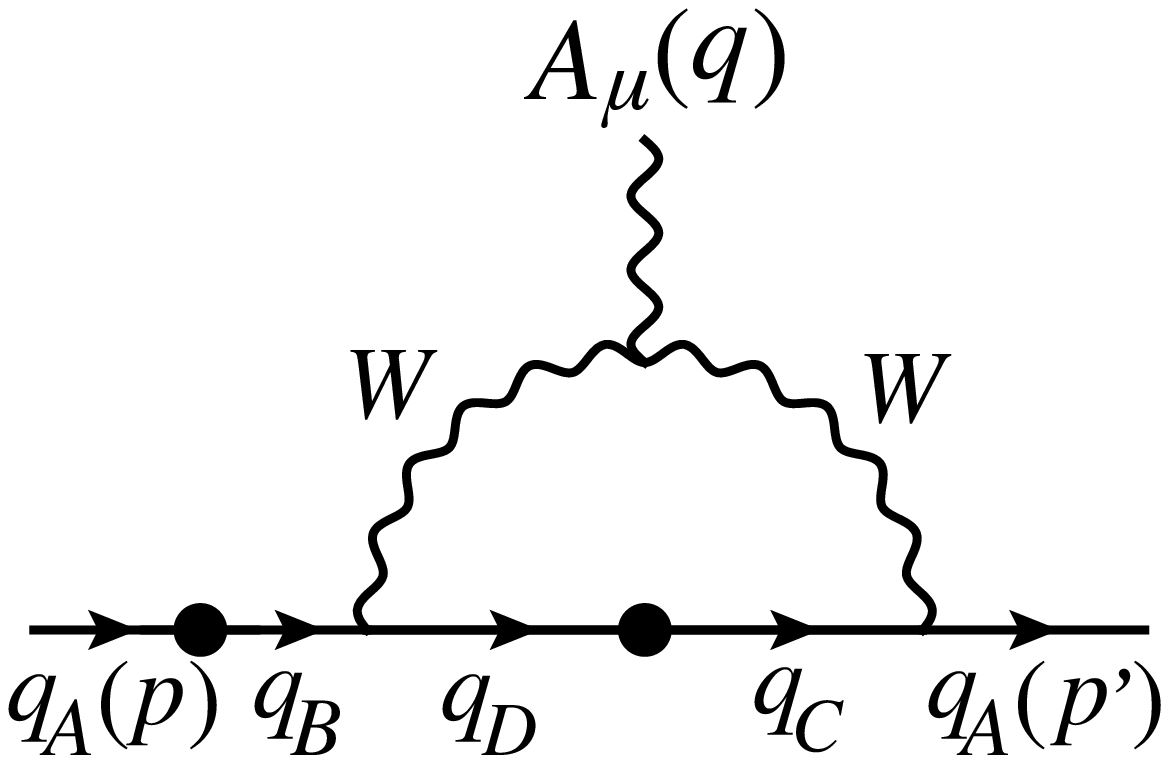}
\caption{\label{wdiags} Feynman diagrams $A_\mu q_Aq_A$ contributing to EMMs, with SME effects entering exclusively through bilinear insertions $q_Aq_B$. These diagrams require the presence of a 3-point vertex $A_\mu W_\alpha W_\beta$ in order to exist.}
\end{figure}

\begin{figure}[ht]
\center
\includegraphics[width=3.2cm]{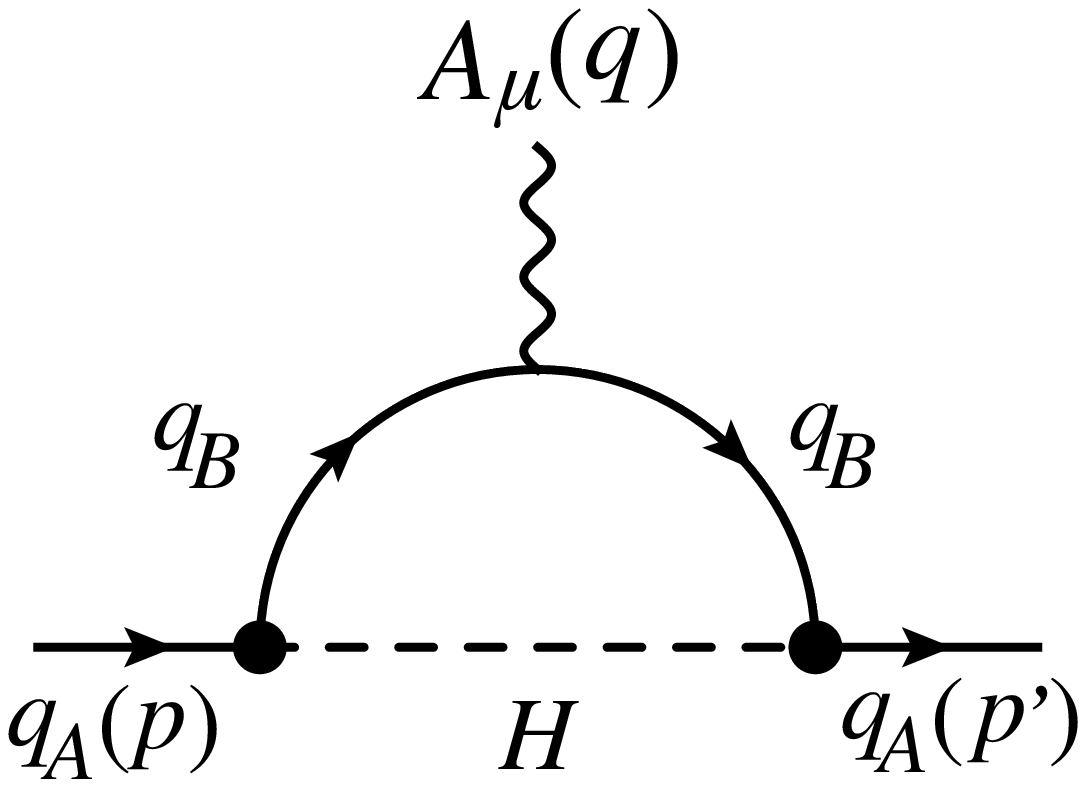}
\hspace{0.3cm}
\includegraphics[width=3.2cm]{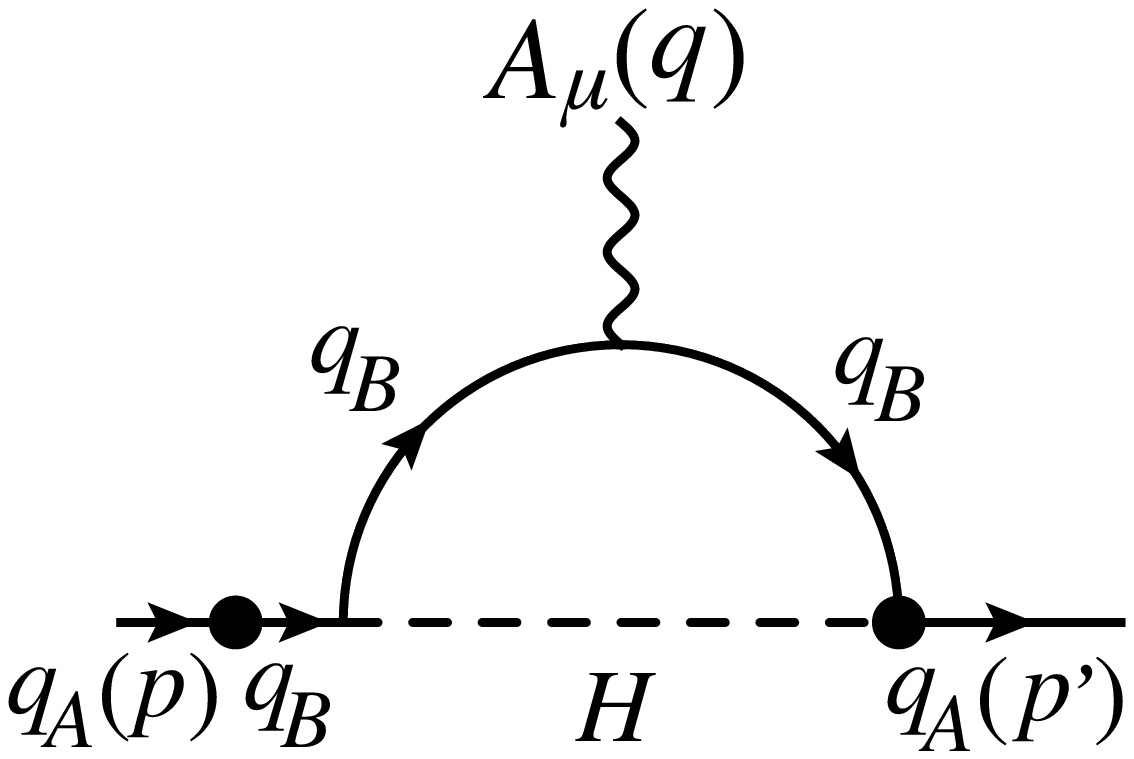}
\includegraphics[width=3.5cm]{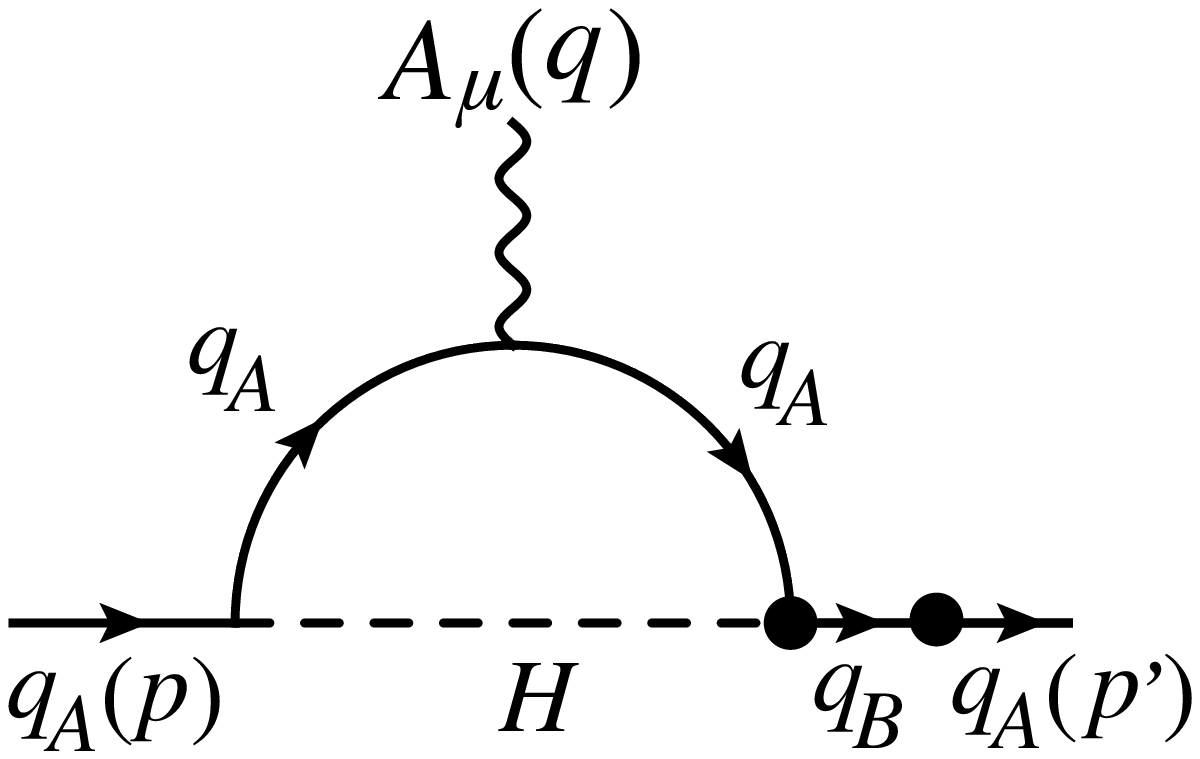}
\hspace{0.3cm}
\includegraphics[width=3.1cm]{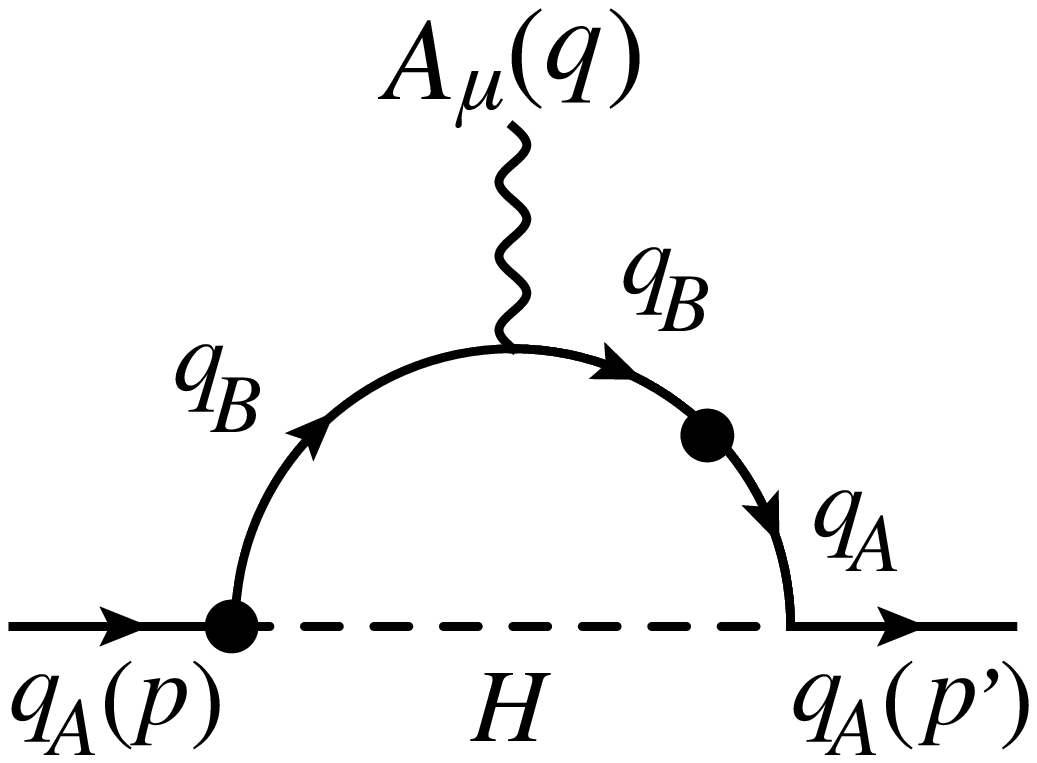}
\includegraphics[width=3.1cm]{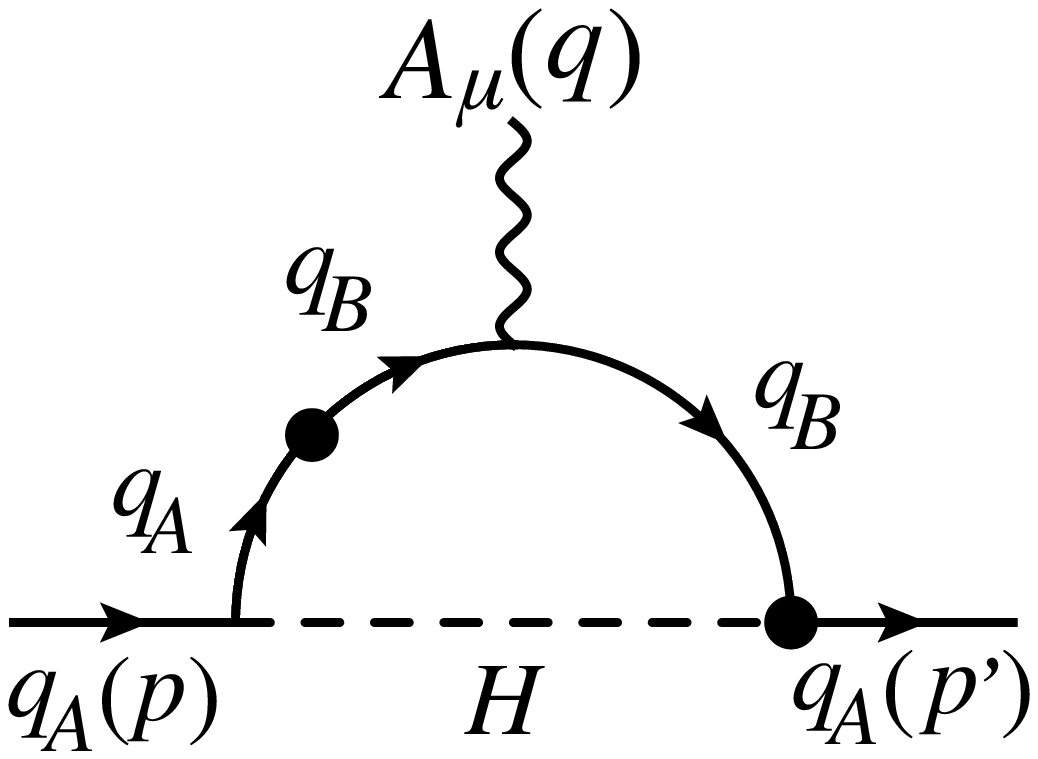}
\hspace{0.3cm}
\includegraphics[width=3.1cm]{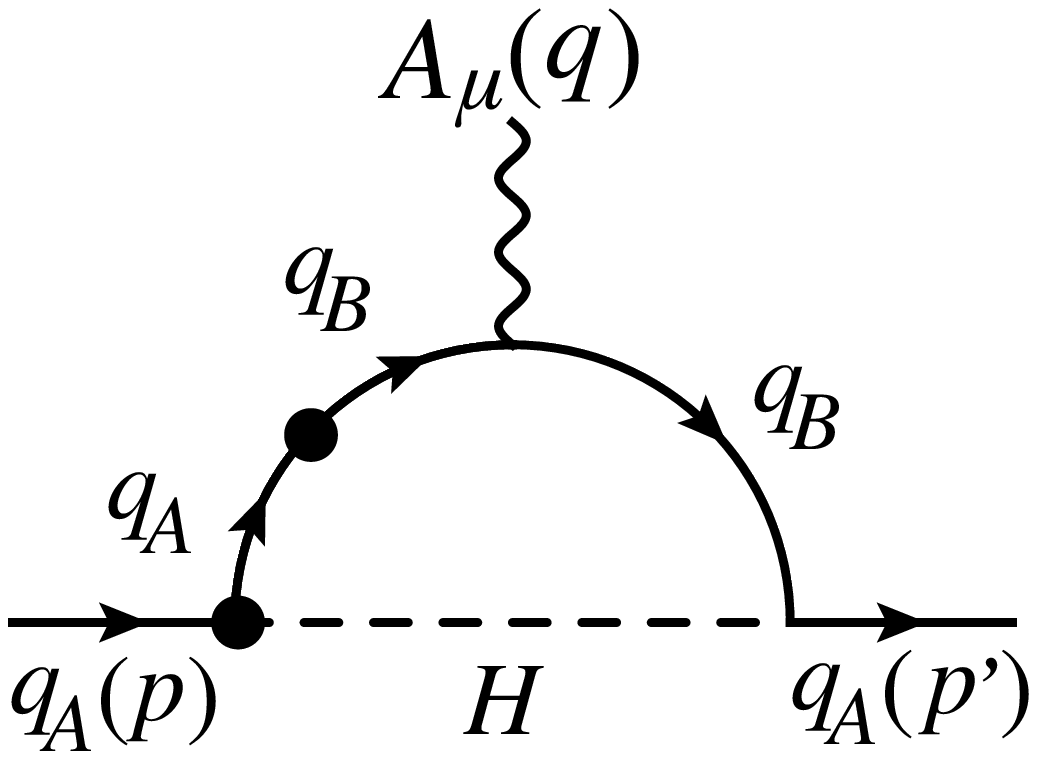}
\includegraphics[width=3.1cm]{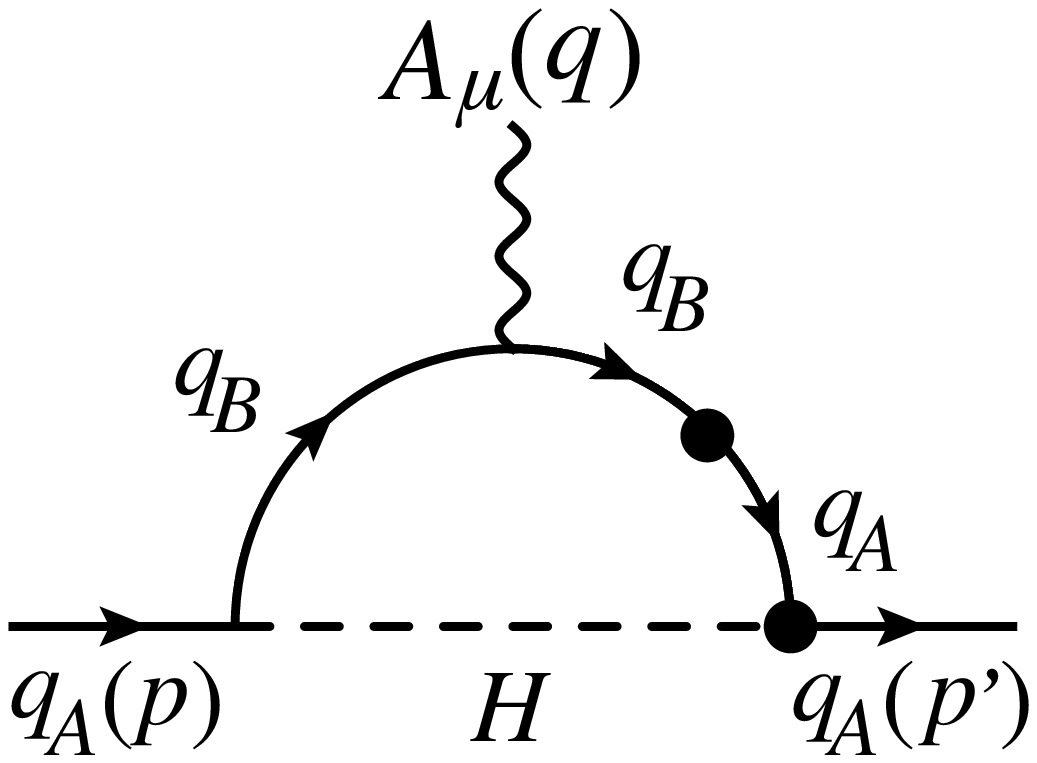}
\hspace{0.3cm}
\includegraphics[width=3.3cm]{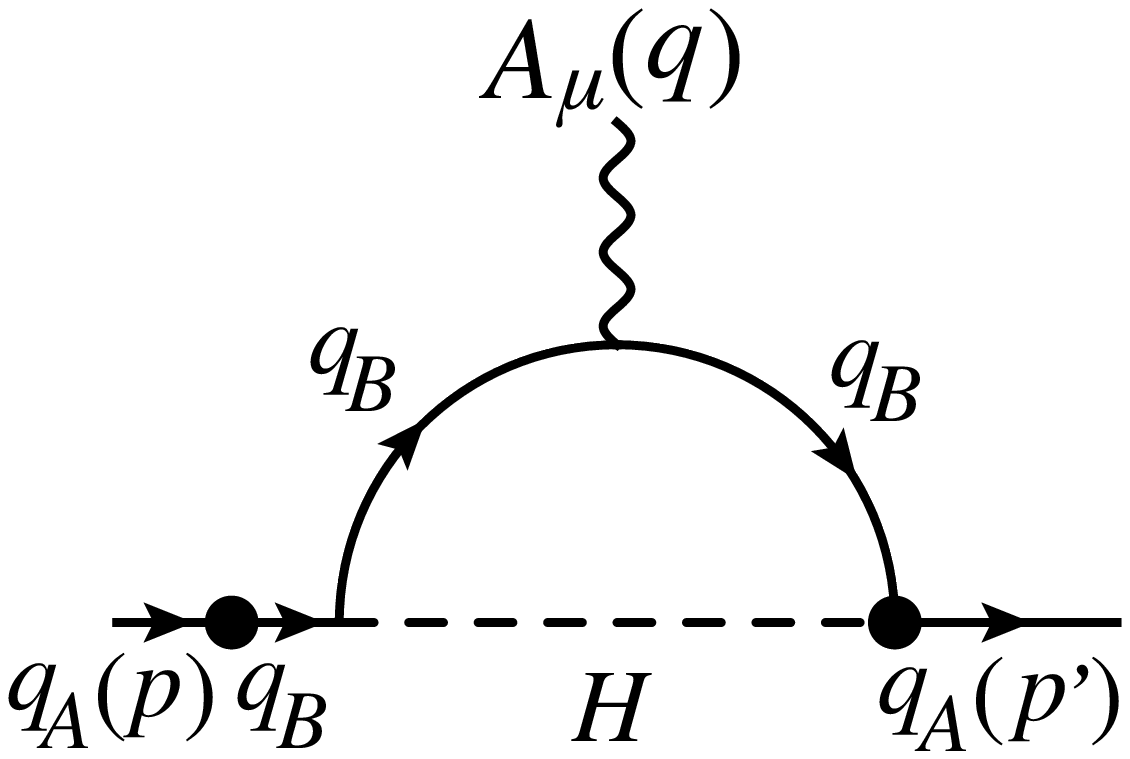}
\caption{\label{HFdiagrams} Feynman diagrams $A_\mu q_A q_A$ contributing to EMMs, with SME effects entering through both bilinear insertions $q_Aq_B$ and three-point vertices $Hq_Aq_B$. These diagrams require the presence of a virtual Higgs-boson line in order to exist.}
\end{figure}

The full set of contributing Feynman diagrams is displayed in Figs.~\ref{gbdiagrams}-\ref{HFdiagrams}. 
The calculation of such diagrams is involved for a number of reasons. Notice that each 2-point insertion comes along with a Dirac propagator, which increases the number of propagator denominators, some of which are subjected to a loop integral. Since two SME insertions, either bilinear, trilinear or both simultaneously, were required in order to generate the Lorentz-invariant EMMs, we found momenta integrals with up to five of such denominators. A detailed discussion on how to proceed can be found in Ref.~\cite{AMNST}. Moreover, the aforementioned extra propagator denominators caused the expressions of the electromagnetic form factors to be gigantic. On the other hand, the resulting amplitude comprises several non-standard tensor structures, in comparison with the conventional Lorentz-invariant parametrization. We spare the reader from dealing with cumbersome enormous expressions and we better omit the explicit results of the full calculation from the paper, concentrating instead in the leading contributions, produced by Schwinger-like diagrams~\cite{Schwinger}. With this in mind, we write the SME contribution to the amplitude $A_\mu q_Aq_A$ as
\begin{eqnarray}
& \displaystyle
\sum_B\big(
\begin{gathered}
\vspace{-0.055cm}
\includegraphics[width=1.16cm]{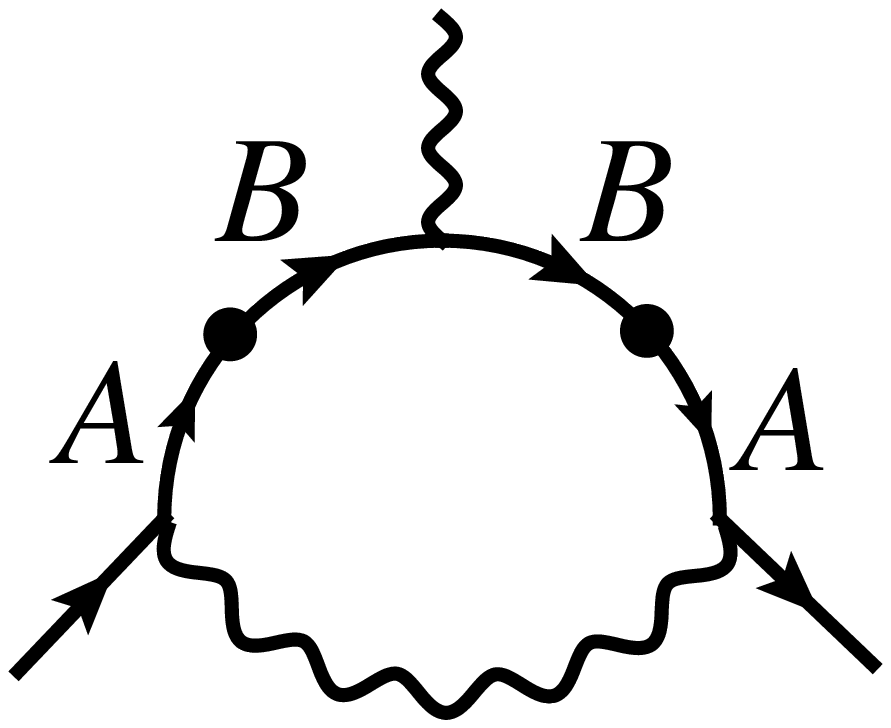}
\end{gathered}
+
\begin{gathered}
\vspace{-0.055cm}
\includegraphics[width=1.16cm]{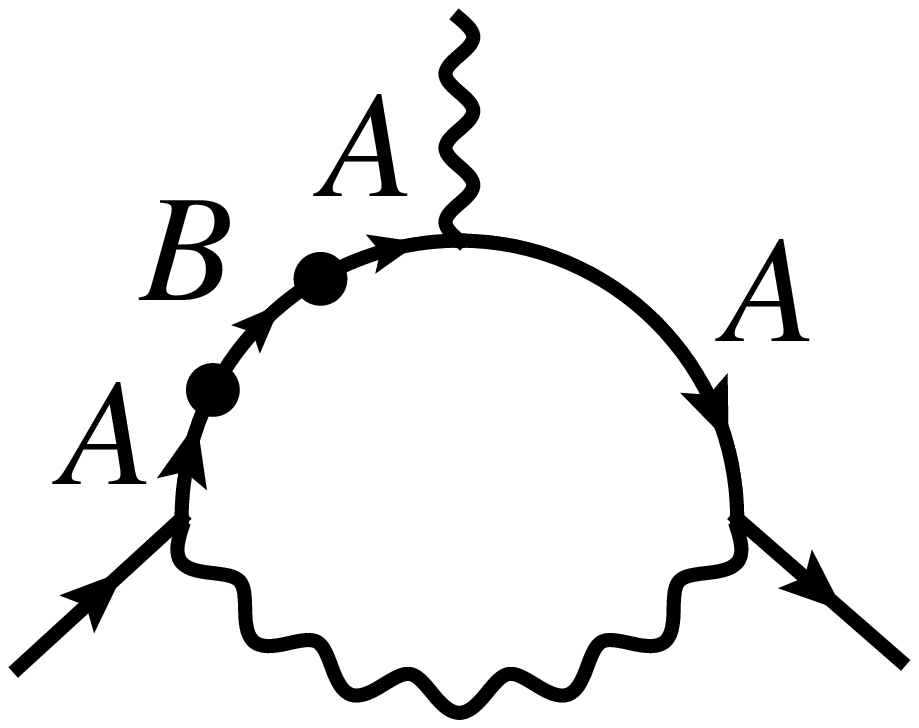}
\end{gathered}
+
\begin{gathered}
\vspace{-0.055cm}
\includegraphics[width=1.16cm]{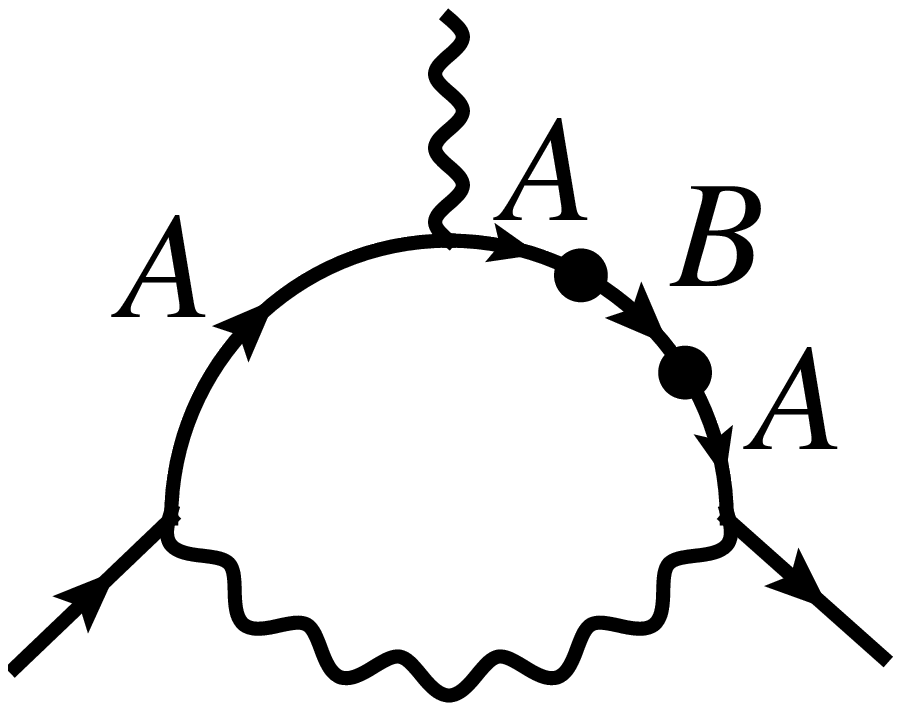}
\label{Aqqstructure}
\end{gathered}
\big)+\,\cdots
\nonumber \\ &
\displaystyle
\hspace{1.3cm}
=\overline{{\mathscr U}_A}\Big(  f^{\rm M}_A\sigma_{\mu\nu}q^\nu+f^{\rm E}_A\gamma_5\sigma_{\mu\nu}q^\nu \Big) {\mathscr U}_A+\cdots,
\end{eqnarray}
where ${\mathscr U}_A$ is a momentum-space quark spinor, and $q$ denotes the incoming momentum of the external photon. Then, $f^{\rm M}_A(q^2)$ and $f^{\rm E}_A(q^2)$ define SME contributions to the AMM $a_A^{\rm SME}$ and the EDM $d_A^{\rm SME}$, respectively, with both quantities preserving Lorentz invariance.  Those diagrams explicitly displayed in the left-hand side of Eq.~(\ref{Aqqstructure}) produce the leading contributions, while the presence of other diagrams has been indicated by the ellipsis. Dots on internal lines of such diagrams represent 2-point insertions, which either preserve or change quark flavor, so $B=A$ or $B\ne A$, with all the corresponding diagrams summed together in Eq.~(\ref{Aqqstructure}). 
Other terms not explicitly shown in the right-hand side of this equation have been indicated by the ellipsis.  Using our calculation of the SME Yukawa-sector contributions to quark electromagnetic form factors, which we identify from Eq.~(\ref{Aqqstructure}), we write the corresponding contributions to AMMs and EDMs as
\begin{eqnarray}
&&
a^{\rm SME}_A=\sum_Ba^{f_A}_B
=\sum_{B}\big[ \tilde{a}_{AB}\big(|{\rm Re}\,{\bf e}^{AB}|^2+|{\rm Re}\,{\bf b}^{AB}|^2\big)
\nonumber \\ &&
\hspace{2.5cm}
+\hat{a}_{AB}\big(|{\rm Im}\,{\bf e}^{AB}|^2+|{\rm Im}\,{\bf b}^{AB}|^2\big) \big],
\label{YSMEa}
\\ \nonumber \\
&&
d^{\rm SME}_A=\sum_B d^{f_A}_B
=\sum_{B}\tilde{d}_{AB}\big( |{\rm Re}\,{\bf e}^{AB}||{\rm Im}\,{\bf b}^{AB}|
\nonumber \\ &&
\hspace{2.5cm}
+|{\rm Re}\,{\bf b}^{AB}||{\rm Im}\,{\bf e}^{AB}| \big).
\label{YSMEd}
\end{eqnarray}
where the sum index $B$ runs over $u,c,t$ if $A$ is a $u$-type quark index, whereas for $A$ being a $d$-type quark index $B=d,s,b$. If $A=B$, the following definitions hold:
\begin{eqnarray}
&&
a^{f_A}_A=\frac{e^3v^2}{(4\pi)^2m_A^2}\Big( 3\big(|{\rm Im}\,{\bf {b}}^{AA}|^2+|{\rm Im}\,{\bf e}^{AA}|^2 \big) 
\nonumber \\ &&
+\big( |{\rm Re}\,{\bf {b}}^{AA}|^2+|{\rm Re}\,{\bf e}^{AA}|^2 \big)\Big(15+8\Big( \Delta_{\rm IR}+\log\frac{\mu^2}{m_A^2} \Big) \Big),
\label{aAA}
\nonumber \\ \\
&&
d_A^{f_A}=\frac{2e^3v^2}{(4\pi)^2m_A^3}\big( |{\rm Re}\,{\bf e}^{AA}||{\rm Im}\,{\bf b}^{AA}|
\nonumber \\ &&
+|{\rm Im}\,{\bf e}^{AA}||{\rm Re}\,{\bf b}^{AA}| \big)\Big( 2\Big( \Delta_{\rm IR}+\log\frac{\mu^2}{m_A^2} \Big)-3 \Big),
\label{dAA}
\end{eqnarray}
whereas for $B\ne A$ we find
\begin{widetext}
\begin{eqnarray}
a^{f_A}_B&=&\frac{e^3v^2}{96\pi^2m_A^2(m_A^2-m_B^2)^2}\Big( \big(|{\rm Im}\,{\bf b}^{AB}|^2+|{\rm Im}\,{\bf e}^{AB}|^2\big) \Big(2 m_A^7\left(12 m_B-17 m_A\right) \log\left(\frac{m_A^2}{m_B^2}\right)
\nonumber \\ &&   
+\left(m^2_A-m^2_B\right)(-24 m_A^5 m_B+13 m_A^4m_B^2-18 m_A^2 m_B^4-2 \left(m_A-m_B\right){}^2\left(m_A+m_B\right) (-7 m_A^2 m_B
\nonumber \\ &&
-9 m_A m_B^2+17m_A^3-9 m_B^3) \log\left(\frac{m_B^2}{m_B^2-m_A^2}\right)+39m_A^6)\Big)-\big( |{\rm Re}\,{\bf b}^{AB}|^2+|{\rm Re}\,{\bf e}^{AB}|^2 \big)(2 m_A^7(17 m_A
\nonumber \\ &&
+12 m_B) \log\left(\frac{m_A^2}{m_B^2}\right)+\left(m^2_A-m^2_B\right) (-24 m_A^5 m_B-13 m_A^4m_B^2+18 m_A^2 m_B^4
\nonumber \\ &&
+2 \left(m_A-m_B\right)\left(m_A+m_B\right){}^2 \left(7 m_A^2 m_B-9 m_A m_B^2+17m_A^3+9 m_B^3\right) \log\left(\frac{m_B^2}{m_B^2-m_A^2}\right)-39m_A^6)) \Big),
\label{aAB}
\\
d_B^{f_A}&=&\frac{e^3v^2m_B^3}{2\pi^2m_A^4(m_B^2-m_A^2)}\big( |{\rm Re}\,{\bf e}^{AB}||{\rm Im}\,{\bf b}^{AB}|+|{\rm Im}\,{\bf e}^{AB}||{\rm Re}\,{\bf b}^{AB}| \big)\log\frac{m_B^2}{m_B^2-m_A^2}.
\label{dAB}
\end{eqnarray}
\end{widetext}
Mass-dependent coefficients $\tilde{a}_{AB}$ and $\hat{a}_{AB}$, which are part of Eq~(\ref{YSMEa}), are derived from Eqs.~(\ref{aAA}) and (\ref{aAB}). Meanwhile, coefficients $\tilde{d}_{AB}$ and $\hat{d}_{AB}$, in Eq.~(\ref{YSMEd}), follow from Eqs.~(\ref{dAA}) and (\ref{dAB}). Since these EMMs contributions are entirely given by real and imaginary parts of Lorentz-violation coefficients ${\bf e}^{AB}$ and ${\bf b}^{AB}$, such quantities are the ones to be compared with experimental results and thus the ones to bound, but keep in mind that the original SME coefficients, which constitute them, are $(H_u)^{AB}_{\mu\nu}$ and $(H_d)^{AB}_{\mu\nu}$, as introduced in Eq.~(\ref{firstLVYukawa}). Another point to remark is the presence of infrared divergences, $\Delta_{\rm IR}$, which we discuss in a moment. \\

While Eqs.~(\ref{YSMEa})-(\ref{dAB}) display explicitly the leading contributions, which correspond to a subset of virtual-photon diagrams, let us comment that we performed the full calculation indeed, which includes diagrams with (1) virtual $W$ bosons, (2) virtual $Z$ bosons, and (3) a virtual Higgs boson. In particular, some of such contributions bear gauge dependence, caused by gauge-boson propagators. For practical reasons, we executed the calculation of these contributions in the unitary gauge, which eliminates pseudo-Goldstone bosons, thus reducing the number of contributing diagrams. A latent issue entailed by this gauge choice originates in the form acquired by gauge-boson propagators, which increases the superficial degree of freedom of loop integrals, thus entangling the elimination of UV divergences from amplitudes that are, presumably, UV finite. However, after an explicit calculation, we have arrived at the conclusion that all our expressions are free of UV divegences. On the other hand, IR divergences did not vanish from these new-physics contributions, though, as argued in Ref~\cite{AMNST}, they are expected to vanish from cross sections. Note that, among the whole set of contributing diagrams, which can be looked at in Figs.~\ref{gbdiagrams}-\ref{HFdiagrams}, those having a virtual photon and involving two 2-point insertions on a single loop fermion line are the only ones giving rise to IR divergences. In order to discuss how the elimination of these IR divergences is expected to happen, let us consider, in the context of Lorentz-invariant quantum electrodynamics, the one-loop contribution to the electromagnetic vertex $A_\mu l_Al_A$, 
\begin{equation}
\begin{gathered}
\vspace{0.15cm}
\includegraphics[width=2cm]{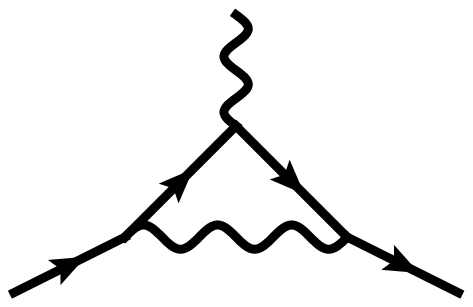}
\end{gathered}
=ie\Big[ \gamma_\mu F_1(q^2)+i\sigma_{\mu\nu}q^\nu\frac{F_2(q^2)}{2m_l} \Big],
\label{QEDemvpar}
\end{equation}
given in terms of two form factors, $F_1(q^2)$ and $F_2(q^2)$, both depending only on the squared incoming external-photon momentum $q^2$ and the lepton mass $m_l$. While the magnetic form factor, $F_2(q^2)$, is free of IR divergences, and from UV divergences as well, the charge form factor, $F_1(q^2)$, bears both sorts of divergences, with the latter eliminated from the amplitude through renormalization. Conversely, IR divergences nested within $F_1(q^2)$ remain in the amplitude, to be later cancelled from some cross section. To illustrate this, think about the total contribution from Lorentz-invariant quantum electrodynamics to the process $l_A^+l_A^-\to l_B^+l_B^-$ at one loop, whose amplitude can be written as ${\cal M}_{2\to2}={\cal M}^{\rm tree}_{2\to2}+{\cal M}^{\rm 1-loop}_{2\to2}$, where ${\cal M}^{\rm tree}_{2\to2}$ is the tree-level contribution. The one-loop term ${\cal M}^{\rm 1-loop}_{2\to2}$, on the other hand, is expressed as
${\cal M}^{\rm 1-loop}_{2\to2}={\cal M}^{\gamma\,l_Al_A}_{2\to2}+\cdots$, where
\begin{equation}
{\cal M}^{\gamma\,l_Al_A}_{2\to2}=
\begin{gathered}
\vspace{-0.1cm}
\includegraphics[width=1.5cm]{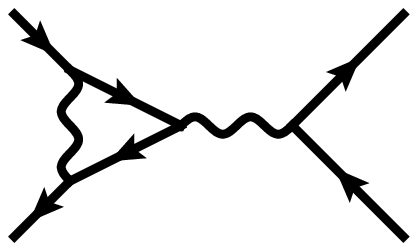}
\end{gathered}
\,\,+\,\,
\begin{gathered}
\vspace{-0.1cm}
\includegraphics[width=1.5cm]{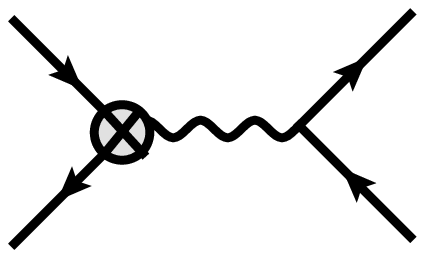}
\end{gathered}
\label{lltollMloopdiags}
\end{equation}
carries the full contribution from the electromagnetic vertex $A_\mu l_Al_A$, as parametrized in Eq.~(\ref{QEDemvpar}), to the amplitude of $l_Al_A\to l_Bl_B$, with the second diagram representing the counterterm, entering as part of the renormalization procedure. Since the interference terms of the differential cross section, given as $d\sigma_{2\to2}^{\rm interf.}\varpropto\sum_{\rm spin}\big[ ({\cal M}^{\rm tree}_{2\to2})^*{\cal M}^{\gamma\,l_Al_A}_{2\to2}+{\cal M}^{\rm tree}_{2\to2}({\cal M}^{\gamma\,l_Al_A}_{2\to2})^* \big]$,  are still IR divergent, they are not to be understood as observables. Nevertheless, the bremsstrahlung process $l_Al_A\to A_\mu l_Bl_B$, with the assumption of soft-photon emission, results in the amplitude ${\cal M}^{\rm tree}_{2\to3}={\cal M}^{\gamma\,l_Al_A}_{2\to3}+\cdots$, where
\begin{eqnarray}
{\cal M}^{\gamma\,l_Al_A}_{2\to3}&=&
\begin{gathered}
\vspace{-0.055cm}
\includegraphics[width=1.5cm]{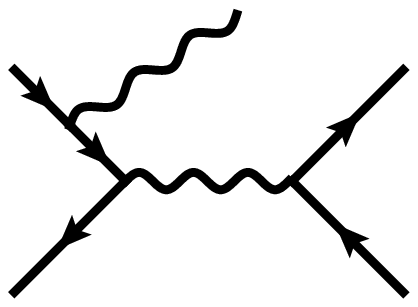}
\end{gathered}
\,\,+\,\,
\begin{gathered}
\vspace{-0.44cm}
\includegraphics[width=1.5cm]{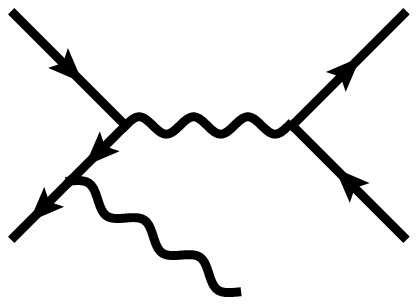}
\label{treelltoll}
\end{gathered}
\,\,.
\end{eqnarray}
The differential cross section of this process, given as $d\sigma_{2\to3}=d\sigma_{2\to3}^{\gamma\,l_Al_A}+\cdots$, with $d\sigma_{2\to3}^{\gamma\,l_Al_A}\varpropto\sum_{\rm spin}|{\cal M}^{\gamma\,l_Al_A}_{2\to3}|^2$, turns out to be IR divergent. Now consider the afore-discussed IR-divergent differential cross sections, $d\sigma^{\rm interf.}_{2\to2}$ and $d\sigma^{\gamma l_Al_A}_{2\to3}$, simultaneously, that is, think about the sum $d\sigma^{\rm interf.}_{2\to2}+d\sigma_{2\to3}^{\gamma\,l_Al_A}$. IR divergences of individual contributing terms of this sum remarkably cancel each other when added together. Regarding this result, we find it worth commenting that any bremsstrahlung diagram of Eq.~(\ref{treelltoll}) is constructed by insertion of the electromagnetic vertex in an external line of some tree-level diagram $l_Al_A\to l_Bl_B$, thus introducing a matrix factor $\gamma_\mu$, which coincides with the one characterizing the IR-divergent form factor $F_1(q^2)$, in Eq.~(\ref{QEDemvpar}).\\ 

With this discussion in mind, let us go back to the IR-divergent contributions from the minimal-SME Yukawa sector  to the electromagnetic vertex $A_\mu q_Aq_A$. The corresponding IR divergences do not arise from the charge form factor, in contrast with the SM case, but they are nested within the electromagnetic form factors instead, meaning that the Dirac-matrix factors going together with them are $\sigma_{\mu\nu}$ and $\sigma_{\mu\nu}\gamma_5$, rather than $\gamma_\mu$. Therefore, bremsstrahlung diagrams with SME bilinear insertions, coming along with matrix factors $\sigma_{\mu\nu}$ and $\sigma_{\mu\nu}\gamma_5$, are reasonably expected to eliminate such divergences when considering soft-photon emission and summing them together with the IR-divergent loop contributions. Aiming at an illustration of this, consider the amplitude characterizing the contributions from the minimal-SME Yukawa sector to $q_Aq_A\to q_Bq_B$, which includes the combination of Feynman diagrams
\begin{equation}
\begin{gathered}
\vspace{-0.055cm}
\includegraphics[width=2.3cm]{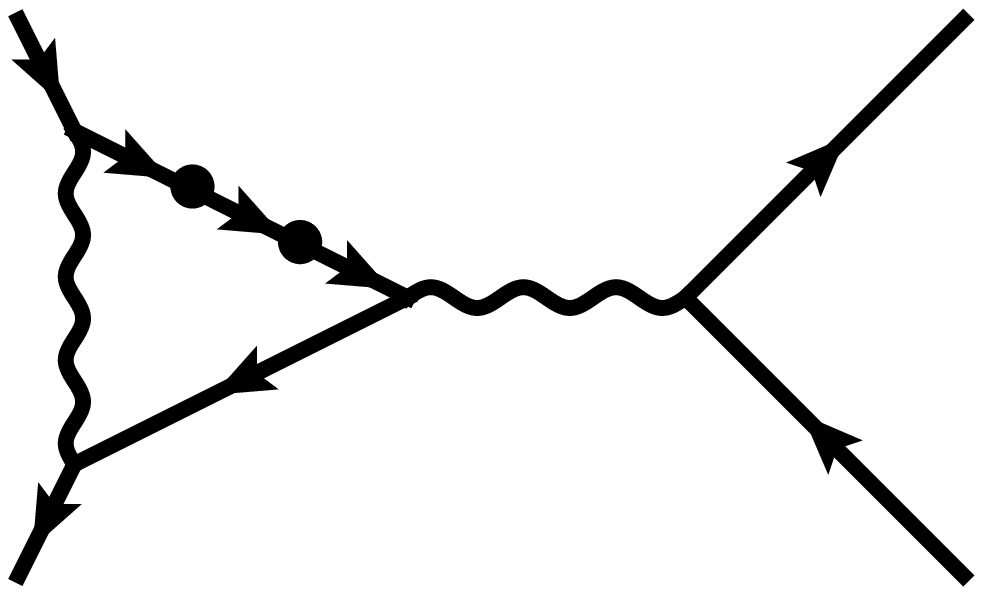}
\end{gathered}
\,\,\,+\,\,\,
\begin{gathered}
\vspace{-0.055cm}
\includegraphics[width=2.3cm]{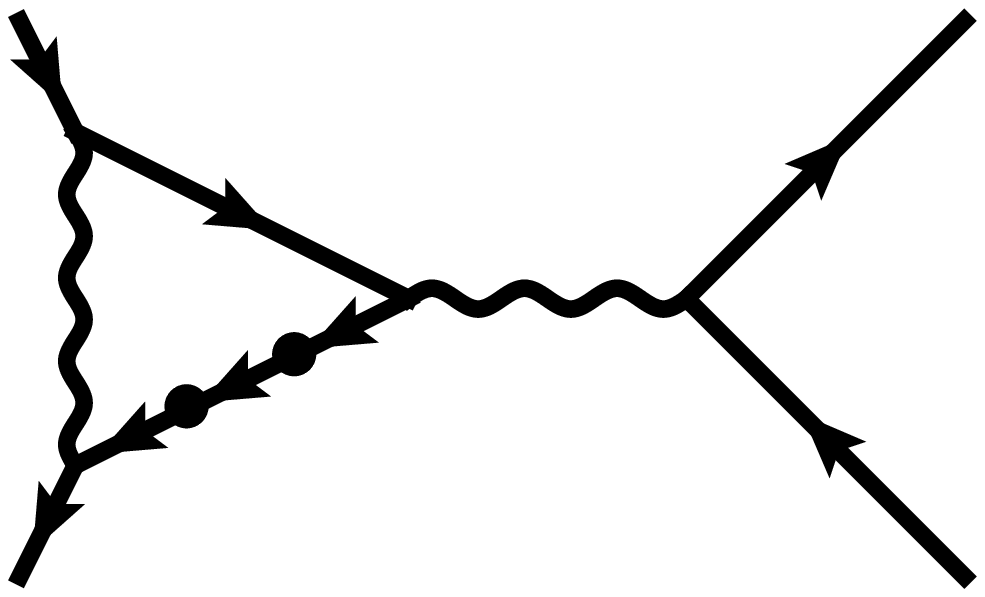}
\end{gathered}
\,\,,
\end{equation}
both involving SME bilinear insertions, at the second order in Lorentz violation, in loop lines. On the other hand, consider the following tree-level diagrams, contributing to the bremsstrahlung process $q_Aq_A\to A_\mu q_Bq_B$:
\begin{equation}
\begin{gathered}
\vspace{-0.055cm}
\includegraphics[width=2.3cm]{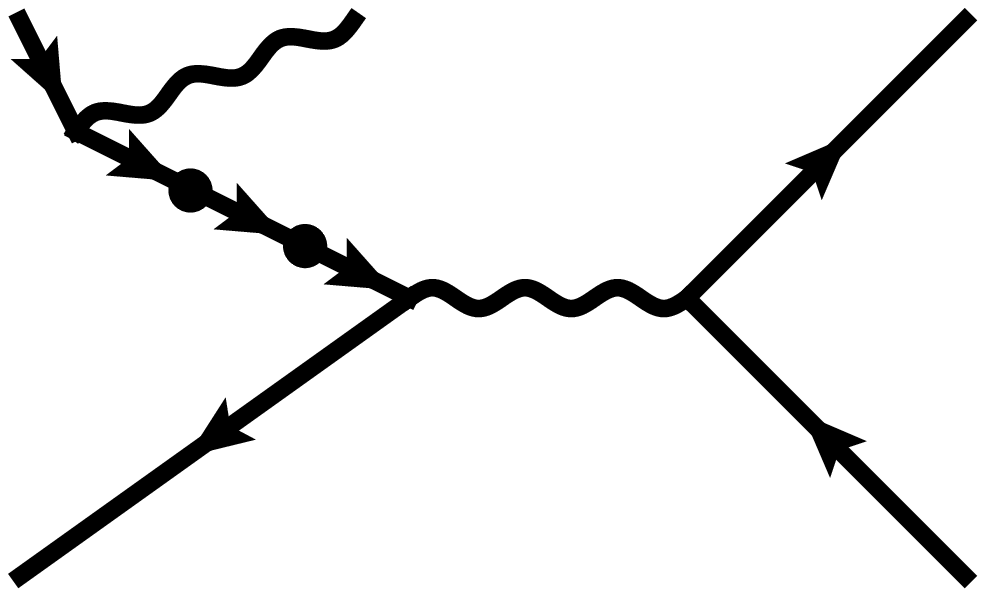}
\end{gathered}
\,\,\,+\,\,\,
\begin{gathered}
\vspace{-0.055cm}
\includegraphics[width=2.3cm]{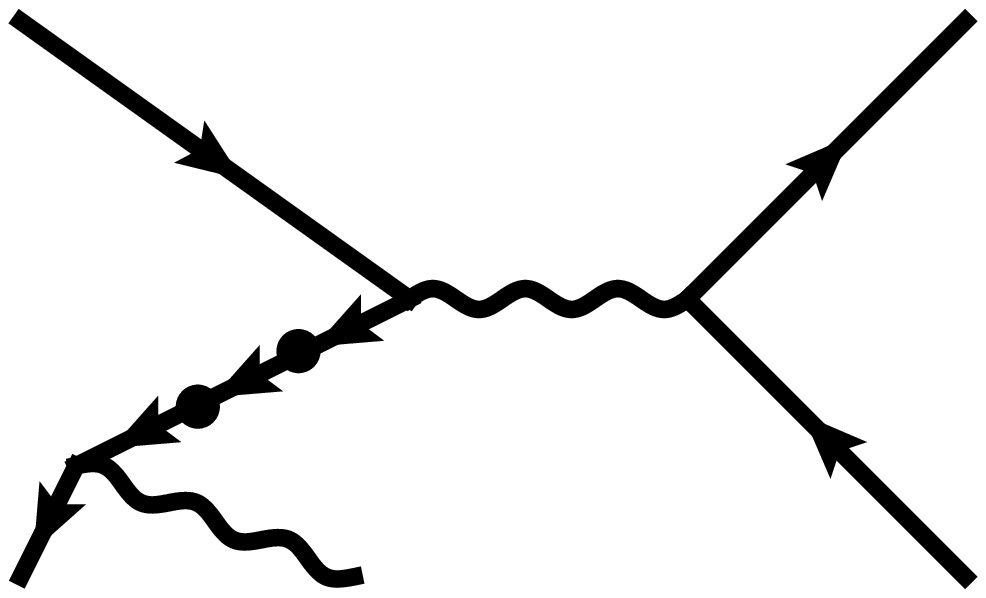}
\end{gathered}
\,\,,
\end{equation}
with the final-state photon assumed to be soft. These tree-level diagrams also include SME bilinear insertions, two of them in a single fermion propagator. Both sums of diagrams, being IR divergent and proportional to matrices $\sigma_{\mu\nu}$ and $\sigma_{\mu\nu}\gamma_5$, are expected to produce contributions to differential cross sections which, summed together, should cancel all IR divergencies, thus resulting in a finite total cross section. The proof of such a statement is not an objective of the present investigation, but a matter of a future work, so, proceeding in a practical manner, we assume that such a cancellation of IR divergences occurs and we ignore them accordingly. Note that the presence of IR divergences within the electromagnetic form factors, at the level of amplitude, implies that the calculated contributions to AMMs and EDMs cannot be considered as observables, though keep in mind that such quantities, presumably finite at the level of cross section, carry contributions to physical processes, which endows them with physical significance. For that reason, their estimation is relevant and well motivated.\\

In general, the AMM and EDM contributions calculated for the present investigation are complex quantities, even though all the external lines have been taken on shell and the one-loop electromagnetic vertex has been assumed to be quark-flavor conserving. The emergence of imaginary parts takes place as long as perturbative 2-point insertions connect external lines to virtual lines of lighter particles, since this induces thresholds. The occurrence of such thresholds can be appreciated in Eqs.~(\ref{aAB}) and (\ref{dAB}), which involve logarithm factors carrying factors $m_B^2-m_A^2$ in their arguments. Recalling that $m_A$ is the mass of the external quark and $m_B$ corresponds to the mass of some virtual quark, note that such differences of quadratic masses can be either positive or negative, thus opening the possibility of having complex quantities. For instance, this happens if $A=t$ and $B=u$, because $m_t>m_u$. An illustrative discussion on  this issue can be found in Ref.~\cite{AMNST}, for the case of leptons, where a set of graphs is provided in which the comportment of real and imaginary parts of contributions to lepton EMMs for different virtual-particle masses $m_B$ is shown. Among the whole set of quark EMMs, the only cases in which the contributions are real correspond to the up and down quarks, for in such cases external lines correspond to particles that are always lighter than virtual particles, thus ensuring the emergence of real-valued contributions. This observation is pertinent to the present investigation, provided that the EMMs of nucleons are calculated from the SME Yukawa-sector contributions to the up and down quark AMMs and EDMs, which are then ensured to be real. \\

The minimal-SME $CPT$-even  sectors of quarks and the Higgs boson, defined in Ref~\cite{CoKo2}, bear couplings which may also produce contributions to quark EMMs. In the case of the quark Lagrangian terms, such contributions would arise from the antisymmetric parts of SME coefficients $c_{\mu\nu}$, which, however, are removed from the theory by an appropriate redefinition of spinor fields~\cite{CoMc}. On the other hand, the contributions from the aforementioned Lorentz-violating Higgs sector carry a suppressing factor $\frac{1}{m_W^4}$, with respect to the contributions calculated in the present investigation, so we disregard them. Finally, let us point out that, in general, the SME Lagrangian terms, being part of an effective field theory, may not all be produced by the genuine fundamental physical description.\\

\section{Bounds on SME coefficients}
\label{estimations}
This section is devoted to the implementation of the analytical expressions previously found, for the leading contributions to quark AMMs and EDMs, to the particular cases of the up and down quarks, aiming at the definition of Lorentz-violation contributions to EMMs of the proton and the neutron. Experimental data on such quantities are then utilized to set bounds on SME coefficients. For the determination of constraints on SME coefficients, we use those values of nucleon EMMs which have been recommended by the Particle Data Group~\cite{PDG}. Relying on a variant of the Penning-Trap experiment, the authors of Ref.~\cite{exppMM} reached the value
\begin{equation}
\mu_p=2.7928473446(8)\mu_N, 
\end{equation}
for the magnetic moment of the proton, given in units of the nuclear magneton, $\mu_N=\frac{e}{2m_p}$, with $m_p$ the mass of the proton. On the other hand, the Committee on Data for Science and Technology, or CODATA for short, published, in Ref.~\cite{expnMM}, a comprehensive list of physical constants, which included the value
\begin{equation}
\mu_n=-1.9130427(5)\mu_N,
\end{equation}
for the neutron magnetic moment. By assuming that new-physics traces might be as large as the errors in these data, the proton magnetic moment measurement, whose error is smaller than the one for the magnetic moment of the neutron by three orders of magnitude, is our choice to constrain SME coefficients. Ref.~\cite{exppDM} provides an estimation of upper bounds on the EDMs of the proton and the neutron, based on the measurement of the EDM of $^{199}{\rm Hg}$ atom, claiming that
\begin{equation}
|d_p|<2.1\times10^{-25}e\cdot{\rm cm},
\end{equation}
in the case of the proton and $|d_n|<2.2\times10^{-26}e\cdot{\rm cm}$ for the neutron. A little improvement on the neutron EDM bound was reported two years later in Ref.~\cite{expnDM}, according to which
\begin{equation}
|d_n|<1.8\times10^{-26}e\cdot{\rm cm},
\end{equation}
reached by using ultracold neutrons and oscillating magnetic fields. Being better by one order of magnitude, the neutron EDM is the one to be used in the present paper to bound SME coefficients.\\

We connect the nucleon EMMs to those of constituent quarks by using the standard prescriptions~\cite{CzKr,MaQue}
\begin{equation}
a_p=\frac{4}{3}a_u-\frac{1}{3}a_d, \hspace{0.3cm} d_n=\frac{4}{3}d_d-\frac{1}{3}d_u.
\label{nucleonEMMs}
\end{equation}
From our calculation of the quark AMM and EDM given in Eqs.~(\ref{YSMEa})-(\ref{dAB}), we write the contribution to the proton magnetic moment and the contribution to the EDM of the neutron as
\begin{eqnarray}
a^{\rm SME}_p&=&\sum_{q=u,d}\sum_{B}\xi^q_p\big[ \tilde{a}_{qB}\big(|{\rm Re}\,{\bf e}^{qB}|^2+|{\rm Re}\,{\bf b}^{qB}|^2\big)
\nonumber \\ &&
\hspace{1.2cm}
+\hat{a}_{qB}\big(|{\rm Im}\,{\bf e}^{qB}|^2+|{\rm Im}\,{\bf b}^{qB}|^2\big) \big],
\label{SMEap}
\\ \nonumber \\
d^{\rm SME}_n&=&\sum_{q=u,d}\sum_{B}\xi^q_n\,\tilde{d}_{qB}\big( |{\rm Re}\,{\bf e}^{qB}||{\rm Im}\,{\bf b}^{qB}|
\nonumber \\ &&
\hspace{1.2cm}
+|{\rm Re}\,{\bf b}^{qB}||{\rm Im}\,{\bf e}^{qB}| \big),
\label{SMEdn}
\end{eqnarray}
respectively. In these expressions, $\xi^u_p=\frac{4}{3}$, $\xi^d_p=-\frac{1}{3}$, $\xi^u_n=-\frac{1}{3}$, and $\xi^d_n=\frac{4}{3}$.
Note, however, that other approaches are available, as, for instance, the one followed in Ref.~\cite{AlSch}, where chiral perturbation theory is utilized to write down Lorentz-violating Lagrangian terms, at the level of hadrons. In the nonperturbative regime of quantum chromodynamics, the SM leading contributions to up- and down-quark AMMs are generated by Schwinger-type Feynman diagrams, amounting to $a^{\rm SM}_u\approx\frac{2\alpha}{9\pi}$ and $a^{\rm SM}_d\approx\frac{\alpha}{18\pi}$, respectively, with $\alpha$ the fine-structure constant. According to Eq.~(\ref{nucleonEMMs}), this yields $a^{\rm SM}_p\approx\frac{5\sqrt{\alpha}}{36\pi^2}\mu_N=1.12\times10^{-3}\mu_N$, for the SM contribution to the AMM of the proton. In the context of the SM, the authors of Ref.~\cite{CzKr} calculated EDMs of quarks, claiming that an accidental cancellation of contributions at the two-loop level forces the first EDM contributions to emerge from three-loop Feynman diagrams. Then, by considering the resulting three-loop contributions to the EDMs of the up and down quarks, and using the prescription given in Eq.~(\ref{nucleonEMMs}), they arrived at the conclusion that the neutron EDM is $|d^{\rm SM}_n|\sim10^{-34}\,e\cdot{\rm cm}$.\\

Through our expression for the SME Yukawa-sector contribution to the proton AMM, we write down  Table~\ref{AMMtab}, 
\begin{table}[ht]
\center
\begin{tabular}{cc}
{\bf LVP} & {\bf Bounds}
\\ \hline \hline
$|{\rm Re}\{ {\bf e}^{uu},{\bf b}^{uu} \}|$ & $6.828\times10^{-10}$
\\ 
$|{\rm Im}\{ {\bf e}^{uu},{\bf b}^{uu} \}|$ & $1.526\times10^{-9}$
\\
$|{\rm Re}\{ {\bf e}^{uc},{\bf b}^{uc} \}|$ & $1.921\times10^{-8}$
\\
$|{\rm Im}\{ {\bf e}^{uc},{\bf b}^{uc} \}|$ & $1.925\times10^{-8}$
\\
$|{\rm Re}\{ {\bf e}^{ut},{\bf b}^{ut} \}|$ & $7.156\times10^{-11}$
\\
$|{\rm Im}\{ {\bf e}^{ut},{\bf b}^{ut} \}|$ & $7.156\times10^{-11}$
\\
$|{\rm Re}\{ {\bf e}^{dd},{\bf b}^{dd} \}|$ & $6.828\times10^{-9}$
\\
$|{\rm Im}\{ {\bf e}^{dd},{\bf b}^{dd} \}|$ & $1.526\times10^{-8}$
\\
$|{\rm Re}\{ {\bf e}^{ds},{\bf b}^{ds} \}|$ & $3.267\times10^{-8}$
\\
$|{\rm Im}\{ {\bf e}^{ds},{\bf b}^{ds} \}|$ & $3.441\times10^{-8}$
\\
$|{\rm Re}\{ {\bf e}^{db},{\bf b}^{db} \}|$ & $2.194\times10^{-7}$
\\
$|{\rm Im}\{ {\bf e}^{db},{\bf b}^{db} \}|$ & $2.220\times10^{-7}$
\\ \hline \hline
\end{tabular}
\caption{ \label{AMMtab} Bounds from the proton magnetic moment on Lorentz-violation parameters (LVP) of the minimal-SME Yukawa sector. ${\rm Re}\{ {\bf e}^{AB},{\bf b}^{AB} \}$ denotes both ${\rm Re}\, {\bf e}^{AB}$ and ${\rm Re}\, {\bf b}^{AB}$, and the same applies for imaginary parts.}
\end{table}
which displays upper bounds on maximal attained sensitivities of the real and imaginary parts of ${\bf e}^{AB}$ and ${\bf b}^{AB}$. The most stringent limits, of order $\sim10^{-11}$, are established on the SME parameters ${\bf e}^{ut}$ and ${\bf b}^{ut}$, which link the physics of Lorentz-invariance violation of the $u$ and $t$ quarks. Let us comment on Penning Trap experiments, which is another method to probe Lorentz invariance though EMMs of electrons, protons and their antiparticles~\cite{pMM,BKR,DiKo,LVandDM}. Such devices, aimed at handling stable particles, have the clear advantage that they are sensitive to first-order contributions in SME coefficients. Notice that, while the path followed by us in the present paper requires second-order contributions from SME coefficients, it has given us access to effects of SME coefficients linked to heavy quarks, inaccessible to Penning-Trap experiments. \\

Next, we discuss bounds from the neutron EDM. Consider some fixed quark-flavor indices $q,B$ in the sum defining the EDM contribution in Eq.~(\ref{SMEdn}) and assume that all contributions associated to other combinations of indices vanish. Then bound the remaining terms through experimental data. The resulting constraints, for each fixed $q,B$, are shown in Table~\ref{comparingEMMs},
\begin{table}[ht]
\center
\begin{tabular}{ccc}
{\bf EMM} & {\bf Bounds} & {\bf Best}
\\ \hline \hline 
$a_p$ & $\big|1.963|{\rm Re}\,{\bf e}^{uu}|^2+0.393|{\rm Im}\,{\bf b}^{uu}|^2\big|<9.2\times10^{-19}$ 
\\
$d_n$ & $0.970|{\rm Re}\,{\bf e}^{uu}||{\rm Im}\,{\bf b}^{uu}|<1.8\times10^{-23}$ & \checkmark
\\ \hline
$a_p$ & $\big|1.963|{\rm Re}\,{\bf b}^{uu}|^2+0.393|{\rm Im}\,{\bf e}^{uu}|^2\big|<9.2\times10^{-19}$
\\
$d_n$ & $0.970|{\rm Re}\,{\bf b}^{uu}||{\rm Im}\,{\bf e}^{uu}|<1.8\times10^{-23}$ & \checkmark
\\ \hline
$a_p$ & $\big|2.479|{\rm Re}\,{\bf e}^{uc}|^2-2.468|{\rm Im}\,{\bf b}^{uc}|^2\big|<9.2\times10^{-16}$
\\
$d_n$ & $2.037|{\rm Re}\,{\bf e}^{uc}||{\rm Im}\,{\bf b}^{uc}|<1.8\times10^{-20}$ & \checkmark
\\ \hline
$a_p$ & $\big|2.479|{\rm Re}\,{\bf b}^{uc}|^2-2.468|{\rm Im}\,{\bf e}^{uc}|^2\big|<9.2\times10^{-16}$
\\
$d_n$ & $2.037|{\rm Re}\,{\bf b}^{uc}||{\rm Im}\,{\bf e}^{uc}|<1.8\times10^{-20}$ & \checkmark
\\ \hline
$a_p$ & $\big|1.787|{\rm Re}\,{\bf e}^{ut}|^2+1.787|{\rm Im}\,{\bf b}^{ut}|^2\big|<9.2\times10^{-21}$ & \checkmark
\\
$d_n$ & $1.497|{\rm Re}\,{\bf e}^{ut}||{\rm Im}\,{\bf b}^{ut}|<1.8\times10^{-18}$
\\ \hline
$a_p$ & $\big|1.787|{\rm Re}\,{\bf b}^{ut}|^2+1.787|{\rm Im}\,{\bf e}^{ut}|^2\big|<9.2\times10^{-21}$ & \checkmark
\\
$d_n$ & $1.497|{\rm Re}\,{\bf b}^{ut}||{\rm Im}\,{\bf e}^{ut}|<1.8\times10^{-18}$
\\ \hline
$a_p$ & $\big|1.963|{\rm Re}\,{\bf e}^{dd}|^2+0.393|{\rm Im}\,{\bf b}^{dd}|^2\big|<9.2\times10^{-17}$
\\
$d_n$ & $6.208|{\rm Re}\,{\bf e}^{dd}||{\rm Im}\,{\bf b}^{dd}|<1.8\times10^{-21}$ & \checkmark
\\ \hline
$a_p$ & $\big|1.963|{\rm Re}\,{\bf b}^{dd}|^2+0.393|{\rm Im}\,{\bf e}^{dd}|^2\big|<9.2\times10^{-17}$
\\
$d_n$ & $6.208|{\rm Re}\,{\bf b}^{dd}||{\rm Im}\,{\bf e}^{dd}|<1.8\times10^{-21}$ & \checkmark
\\ \hline
$a_p$ & $\big|-8.575|{\rm Re}\,{\bf e}^{ds}|^2+7.727|{\rm Im}\,{\bf b}^{ds}|^2\big|<9.2\times10^{-15}$
\\
$d_n$ & $4.329|{\rm Re}\,{\bf e}^{ds}||{\rm Im}\,{\bf b}^{ds}|<1.8\times10^{-20}$ & \checkmark
\\ \hline
$a_p$ & $\big|-8.575|{\rm Re}\,{\bf b}^{ds}|^2+7.727|{\rm Im}\,{\bf e}^{ds}|^2\big|<9.2\times10^{-15}$
\\
$d_n$ & $4.329|{\rm Re}\,{\bf b}^{ds}||{\rm Im}\,{\bf e}^{ds}|<1.8\times10^{-20}$ & \checkmark
\\ \hline
$a_p$ & $\big|-1.901|{\rm Re}\,{\bf e}^{db}|^2+1.857|{\rm Im}\,{\bf b}^{db}|^2\big|<9.2\times10^{-14}$
\\
$d_n$ & $0.990|{\rm Re}\,{\bf e}^{db}||{\rm Im}\,{\bf b}^{db}|<1.8\times10^{-19}$ & \checkmark
\\ \hline
$a_p$ & $\big|-1.901|{\rm Re}\,{\bf b}^{db}|^2+1.857|{\rm Im}\,{\bf e}^{db}|^2\big|<9.2\times10^{-14}$
\\
$d_n$ & $0.990|{\rm Re}\,{\bf b}^{db}||{\rm Im}\,{\bf e}^{db}|<1.8\times10^{-19}$ & \checkmark
\\ \hline \hline
\end{tabular}
\caption{\label{comparingEMMs} Comparison of sensitivities to SME coefficients of experimental bounds from the proton magnetic moment and the neutron EDM.}
\end{table}
which is an illustrative mean to compare sensitivities to SME coefficients of experiments that measure proton AMMs with those aimed at neutron EDMs. In most cases the best constraints are given by the neutron EDM experimental bound, being $q=u,B=t$ the only case in which the proton AMM yields the most stringent limit. In this framework, consider a scenario in which all SME parameters $|{\rm Re}\,{\bf e}^{AB}|$, $|{\rm Re}\,{\bf b}^{AB}|$, $|{\rm Im}\,{\bf e}^{AB}|$, and $|{\rm Im}\,{\bf b}^{AB}|$ are practically equal to each other. Under such circumstances, the neutron EDM experimental bound gives rise to Table~\ref{EDMstab},
\begin{table}[ht]
\center
\begin{tabular}{cc}
 {\bf LVP} & {\bf Bounds}
\\ \hline \hline 
$|{\bf e}^{uu},{\bf b}^{uu}|$ & $4.308\times10^{-12}$
\\
$|{\bf e}^{uc},{\bf b}^{uc}|$ & $9.401\times10^{-11}$
\\
$|{\bf e}^{ut},{\bf b}^{ut}|$ & $1.096\times10^{-9}$
\\
$|{\bf e}^{dd},{\bf b}^{dd}|$ & $1.703\times10^{-11}$
\\
$|{\bf e}^{ds},{\bf b}^{ds}|$ & $6.449\times10^{-11}$
\\
$|{\bf e}^{db},{\bf b}^{db}|$ & $4.264\times10^{-10}$
\\ \hline \hline
\end{tabular}
\caption{ \label{EDMstab} Bounds from the neutron electric moment on LVPs of the minimal-SME Yukawa sector. $|{\bf e}^{AB},{\bf b}^{AB}|$ denotes both $|{\bf e}^{AB}|$ and $|{\bf b}^{AB}|$.}
\end{table}
according to which Lorentz-violation parameters $|{\bf e}^{uu}|$ and $|{\bf b}^{uu}|$ are restricted to be as small as $10^{-12}$, thus improving bounds from AMMs (Table~\ref{AMMtab}) by about two orders of magnitude.\\

\section{Summary and conclusions}
\label{conclusions}
The present work comprises a calculation of Lorentz-invariant electromagnetic properties of quarks, developed in the context of the Lorentz- and $CPT$-violating Standard Model Extension, usually referred to by the acronym SME. Being an effective field theory, the SME allows for the analysis and quantification, at current experimental sensitivity, of some fundamental physical description of nature that produces effects not preserving Lorentz symmetry, without even knowing the actual high-energy formulation behind such effects. The calculation has been restricted to the renormalizable part of this Lorentz-nonpreserving effective theory, the so-called minimal SME. More specifically, we have considered the Yukawa sector of this field formulation, which is characterized by Yukawa-like constants endowed with spacetime indices and which are antisymmetric with respect to them. Implementation of the Brout-Englert-Higgs mechanism then defined a Lagrangian density involving couplings that yield 2-point insertions and 3-point vertices, both bearing Lorentz violation and contributing to low-energy amplitudes upon insertion in Feynman diagrams. \\

An intricate calculation of the electromagnetic vertex $A_\mu q_Aq_A$, performed in the unitary gauge, leaded us to analytic expressions for the one-loop contributions to AMMs and to EDMs of quarks. We argued that Lorentz-violating contributions from the Yukawa sector of the SME to the these quantities, defined to be Lorentz invariant, arise for the very first time at the second order in Lorentz violation. Our results are free of UV divergences, but involve IR divergences, meaning that these quantities are not measurable, but can be rendered finite only at the level of cross section and then contribute to some physical process. Since our analytical results are vast, we reported explicitly only the leading contributions to the electromagnetic form factors, generated by Schwinger-like Feynman diagrams. The aforementioned antisymmetry of Lorentz-violating Yukawa-like constants is used to define constant electric-like fields, ${\bf e}^{AB}$, and magnetic-like fields, ${\bf b}^{AB}$, where $A,B$ are quark-flavor indices. The analytic expressions of the contributions from the SME to quark electromagnetic factors are then entirely given in terms of absolute values of the real and imaginary parts of ${\bf e}^{AB}$ and ${\bf b}^{AB}$, which are thus the quantities to be compared with experimental data and then bounded.\\

Once calculated the analytical expressions for the leading contributions to AMMs and the EDMs of quarks, we particularized our generic expressions to the cases of the up and down quarks. This was then utilized to define EMMs of the proton and the neutron, as high-precision measurements of such quantities are available. Regarding the magnetic moment, the one characterizing the proton turned out to be the best option, whereas bounds on the EDM of the neutron, being the most stringent, were used. Then our estimations of the EMMs of nucleons and their comparison with experimental data yielded constraints on SME coefficients which parametrize effects of Lorentz violation at low energies. Bounds on these quantities as restrictive as $10^{-12}$ were established.

\section*{Acknowledgements}
\noindent
We acknowledge financial support from CONACYT and SNI (M\'exico).

\end{document}